\documentclass[]{article}
\usepackage{authblk}

\usepackage[utf8]{inputenc} 
\usepackage[T1]{fontenc}    
\usepackage{hyperref}       
\usepackage{url}            
\usepackage{booktabs}       
\usepackage{amsfonts}       
\usepackage{nicefrac}       
\usepackage{microtype}      
\usepackage{fullpage}

\usepackage{adjustbox}
\usepackage{hyperref}
\usepackage{url}
\usepackage{graphicx}
\usepackage{amsmath}
\usepackage{bbm}
\usepackage{paralist}
\usepackage{Definitions}
\usepackage[ruled,noline]{algorithm2e} 
\usepackage{color}
\usepackage{enumitem}
\usepackage{mathtools}
\usepackage{subfig}
\usepackage{graphicx}
\usepackage{tabularx}
\usepackage[dvipsnames]{xcolor}

\newcommand{\xhdr}[1]{\vspace{1mm}\noindent{{\bf #1. }}}

\newcommand{\badge}[1]{\ifmmode\text{\texttt{#1}}\else\texttt{#1}\fi}

\newcommand{\explain}[2]{\underset{\mathclap{\overset{\uparrow}{#2}}}{#1}}
\newcommand{\explainup}[2]{\overset{\mathclap{\underset{\downarrow}{#2}}}{#1}}
\newcommand*{\defeq}{\coloneqq}

\newcommand\blfootnote[1]{%
  \begingroup
  \renewcommand\thefootnote{}\footnote{#1}%
  \addtocounter{footnote}{-1}%
  \endgroup
}

 \graphicspath{{./figures/}}

\title{Harnessing Natural Experiments to Quantify \\ the Causal Effect of Badges}

\author[1]{Tomasz Kusmierczyk$^{*}$}
\author[2]{Manuel Gomez-Rodriguez}

\affil[1]{Norwegian University of Science and Technology, tomaszku@idi.ntnu.no}
\affil[2]{Max Planck Institute for Software Systems, manuelgr@mpi-sws.org}

\date{}

\begin{document}


\maketitle

\begin{abstract}
A wide variety of online platforms use digital badges to encourage users to take certain types of \emph{desirable} actions. 
However, despite their growing popularity, their \emph{causal effect} on users'{} behavior is not well understood. 
This is partly due to the lack of counterfactual data and the myriad of complex factors that influence users'{} behavior
over time.
As a consequence, their design and deployment lacks general principles.

In this paper, we focus on first-time badges, which are awarded after a user takes a particular type of
action for the first time, and study their causal effect by harnessing the delayed introduction of several 
badges in a popular Q\&A website. 
In doing so, we introduce a novel causal inference framework for first-time badges whose main technical 
innovations are a robust survival-based hypothesis testing procedure, which controls for the 
he\-te\-ro\-ge\-neity in the benefit users obtain from ta\-king an action, and a bootstrap difference-in-differences method, 
which controls for the random fluctuations in users'{} behavior over time.
Our results suggest that first-time badges steer users'{} behavior if the 
initial benefit a user obtains from taking the corresponding action is sufficiently low, otherwise, we do not find 
significant effects.
Moreover, for badges that successfully steered user behavior, we perform a counterfactual analysis and 
show that they significantly improved the functioning of the site at a community level.

\end{abstract}

\blfootnote{$^{*}$\scriptsize This work was done during Tomasz Kusmierczyk'{}s internship at the Max Planck
Institute for Software Systems.}

\section{Introduction} 
\label{sec:intro}
In recent years, social media sites and online communities have increasingly relied on digital badges to reward their users for different types of online behavior. 
Similarly as their physical counterpart, digital badges have been used both as a reputation mechanism, summarizing the skills and accomplishments of the users who receive 
them, and as an \emph{incentive} mechanism, encouraging users to take certain type of \emph{desirable} actions.
The promise of digital badges is that automated fine-grained monitoring and greater degree of control will help refine their design as incentive mechanisms, increasing users'{} 
engagement and improving the functioning of the corres\-ponding online platform.
However, to fulfill this promise, it is necessary to better understand their causal effect on the online behavior of the users who may receive them---identify \emph{when} and 
\emph{why} they are (not) able to steer their behavior.

In this paper, we focus on first-time badges, which are awarded after a user takes a particular type of action for the first time. First-time badges are the simplest 
type of the highly popular threshold ba\-dges~\cite{anderson2013steering, mutter2014behavioral, zhang2016social}, which are awarded after a user has taken 
an action a pre-specified number of times. 
More specifically, we study the causal effect induced by first-time badges by harnessing several \emph{natural experiments} in Stack 
Overflow\footnote{\scriptsize https://stackoverflow.com}, a popular Q\&A website, consisting of the \emph{delayed} introduction of a badge 
some time after the site'{}s inception.
Despite their simplicity, we need to tackle several challenges, which require careful reasoning:
\vspace{1mm}

\emph{--- Measuring progress towards the badge:} since first-time badges are awarded after performing just one single action, the action count does not provide a direct measure of progress 
towards the badge. This is in contrast with (non-binary) threshold badges, which were typically the focus of previous work~\cite{anderson2013steering, mutter2014behavioral, zhang2016social}.

\emph{--- Utility heterogeneity:} 
the benefit (or \emph{utility}) each user obtains from taking an action differs wildly due to, \eg, user'{}s intrinsic motivation, the target of the action, or other users'{} actions. As a consequence, the times users take to perform an action for the first time spans a large range of values. 

\emph{--- Random temporal changes:} one can frequently observe random fluctuations in users'{} behavior over time due to many different complex factors. 
As a consequence, to assess the strength of the causal effect induced by a badge, it is necessary to control for these random fluctuations.

\vspace{1mm} 
\noindent We address the above mentioned challenges by develo\-ping a novel causal inference framework for first-time badges, especially designed for our 
problem setting. 
Our framework avoids modeling the mechanisms underlying individual user actions and instead adopts a data-driven approach based on survival analysis 
and statistical hypothesis testing. At the heart of our approach there are two technical innovations: 
(i) a robust survival-based hypothesis testing procedure, 
which allows us to account for the utility heterogeneity,
and (ii) a \emph{bootstrap difference-in-differences} method, inspired by the economics literature on natural experiments~\cite{lechner2011estimation, meyer1995natural, rosenzweig2000natural}, which allows us to control for 
the random fluctuations in users'{} behavior over time.
Moreover, while our framework focuses on first-time badges, we argue that our methodological innovations will also shed light on more complex 
badges, \eg, non-binary threshold badges.

In contrast with recent empirical studies on threshold ba\-dges~\cite{anderson2013steering, mutter2014behavioral, oktay2010causal, zhang2016social}, which typically 
assume or conclude that, to some degree, badges always steer users'{} behavior, we do not find statistically significant 
evidence to back up this conclusion 
in all the first-time badges we considered. 
Instead, we provide 
empirical evidence of a more subtle picture.
In particular, our results suggest that first-time badges steer users'{} behavior if the utility a user obtains from taking the corresponding action is sufficiently 
low, otherwise, the badge does not seem to have a significant effect.
Moreover, we hypothesize that this may be also the case for non-binary threshold badges and thus argue that the user utilities should be carefully considered on the 
design and deployment of badges.
Finally, for badges that successfully steered user behavior, we go a step further and, using a survival-based counterfactual analysis, show that they significantly improve 
the functioning of the site at a community level.

\xhdr{Related work}
Our work contributes to the growing literature on ba\-dges~\cite{abramovich2013badges, anderson2013steering, FM7299, ImmStoSyr15, zhang2016social, mutter2014behavioral, oktay2010causal, easley2016incentives, Anderson:2014:EMO:2566486.2568042}, which can be broadly divided into theoretical and empirical studies.

Theoretical studies on ba\-dges~\cite{ImmStoSyr15, zhang2016social, easley2016incentives} analyze the effect of badges on users'{} behavior under stylized 
models of badges, which make strong assumptions, often without empirical support.
Moreover, they typically ignore the inherent utility a user receives from taking the action the badge rewards---the action payoff and cost. 
In contrast, in our work, we avoid making strong assumptions about the mechanisms underlying individual user actions and instead adopt a data-driven
approach, which enable us to account for the utility a user obtains from taking an action.

Empirical studies on ba\-dges~\cite{abramovich2013badges, anderson2013steering, FM7299, mutter2014behavioral, oktay2010causal,Anderson:2014:EMO:2566486.2568042} have mainly focused on threshold badges, where the action count provides a direct measure of progress towards the badge. 
In this context, several authors~\cite{mutter2014behavioral, anderson2013steering} have provided empirical evidence in favor of the \emph{goal-gradient hypothesis}, 
which posits users increase their engagement as they get closer to earning a badge, while other authors~\cite{abramovich2013badges, Anderson:2014:EMO:2566486.2568042} found that degree of success of a badge may depend on different complex factors. 
However, most of these studies did not have access to control groups, which would have allowed them to assess users'{} behavior in the absence of a badge and control for 
random fluctuations in users'{} behavior over time.
A notable exception is by Bornfeld et al.~\cite{FM7299}, which has been concurrently and independently conducted with our work, and it also leverages
natural experiments in the context of badges. However,  in contrast to our work, they rely on standard statistical tests on aggregated counts, account for the 
temporal fluctuations in users'{} behavior in an ad-hoc manner, and ignore the utility heterogeneity across users. 

Moreover, badges can be viewed as a \emph{gamification} mechanism, \ie, a mechanism in which game-design elements are used in a non-gaming environment
to encourage users to perform certain tasks ~\cite{DBLP:conf/chi/DeterdingSNOD11,DBLP:journals/ijmms/SeabornF15}.
Therefore, our research adds up to the existing studies on practical implications of gamification for, \eg, increasing user engagement in learning~\cite{DBLP:journals/ce/HanusF15, DBLP:journals/chb/HamariSRCAE16} and e-learning~\cite{DBLP:conf/teem/GeneMB14, Muntean}, shaping healthy behavioral patterns~\cite{DBLP:conf/mm/GobelHWMS10, DBLP:conf/sgda/McCallumB13, johnson2016gamification},
systems design~\cite{herzig2012gamification} or crowdsourcing~\cite{DBLP:conf/hicss/MorschheuserHK16}.

Finally, natural experiments~\cite{meyer1995natural, rosenzweig2000natural}, difference-in-difference designs~\cite{donald2007inference, lechner2011estimation} and propensity score 
matching~\cite{dehejia2002propensity, stuart2010matching} have been increasingly used to identify causal effects from observational data in online settings, \eg, 
social influence~\cite{aral2012identifying, althoff2017online, chen2011online, kramer2014experimental} or network formation~\cite{jacobs2015assembling, phan2015natural}. 
However, together with Bornfeld et al.~\cite{FM7299}, the present work is one of the first that leverage natural experiments to quantify causal effects in the context of badges.

\section{Data description} 
\label{sec:data}

Our Stack Overflow dataset\footnote{Publicly available at https://archive.org/details/stackexchange} comprises of all individual timestamped actions performed by all 
users from the site'{}s inception from July 31, 2008 to September 14, 2014, which allow us to track the complete sequence of actions users take.

\begin{figure}[t]
\centering
\includegraphics[width=0.315\textwidth]{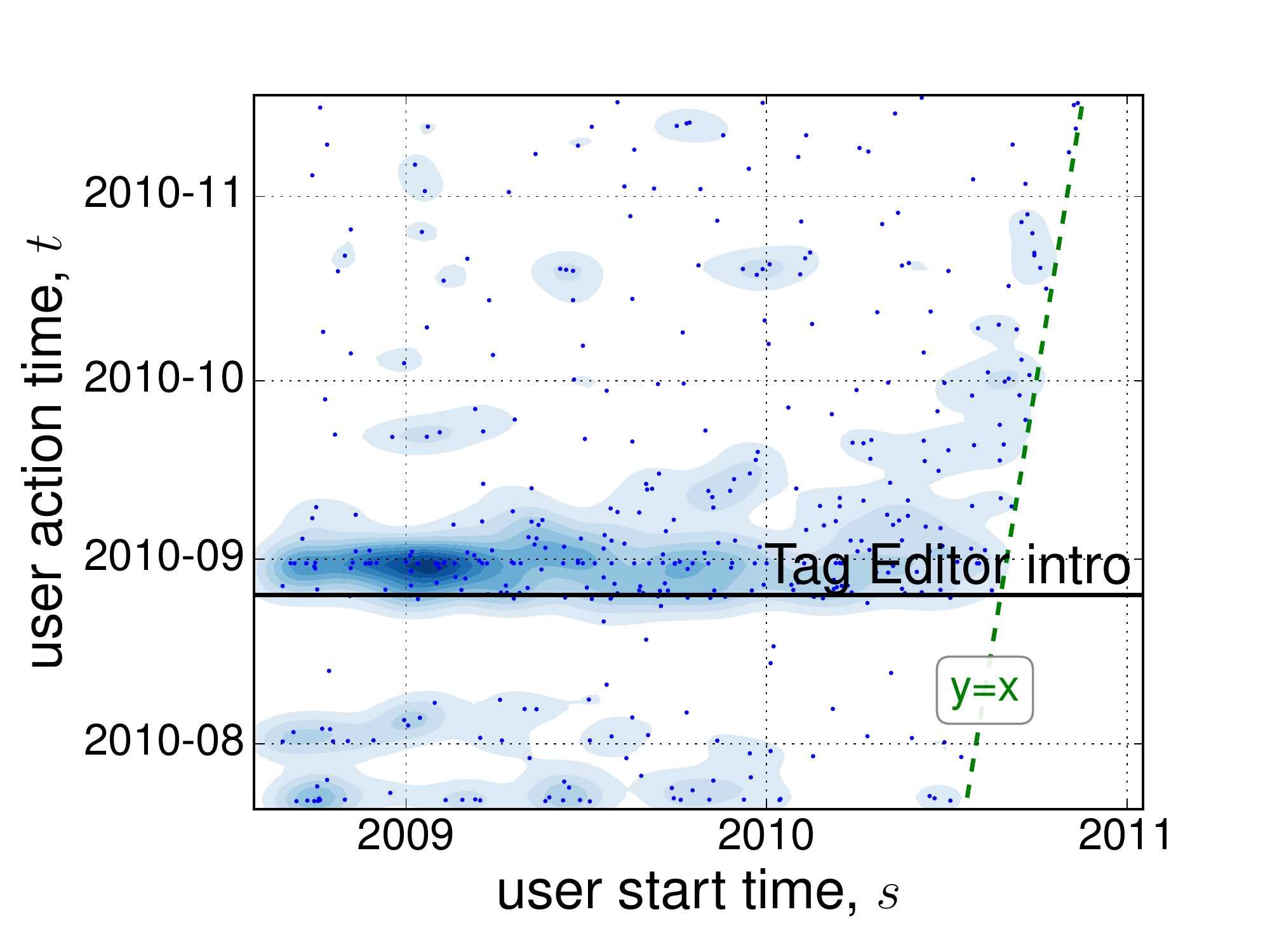}
\vspace{-2mm}
\caption{Time when users first edited a tag wiki (user action time, $t$) against time when they became eligible to edit tag wikis (user start time, $s$). The horizontal black line denotes the  time when 
the \badge{Tag editor} badge was introduced, which is awarded after a user edits a tag wiki for the first time.}
\label{fig:tagedits:points}
\vspace{-3mm}
\end{figure}

\xhdr{First-time badges: natural experiments}
There are a great variety of badges, which reward users for different types of behaviors.
In this work, we focus on \emph{first-time} badges, which are awarded after a user takes a particular type of action for the first time,
and identify those that were introduced some time after the site'{}s inception.
The \emph{delayed} introduction 
of these badges can be thought of as \emph{natural experiments}~\cite{meyer1995natural, rosenzweig2000natural}, where the treatment of
earning a badge is assigned \emph{as if random} among users.
Figure~\ref{fig:tagedits:points} illustrates an example of such badge.
 
More specifically, we select three first-time badges that reward actions whose utilities to the users are clearly different:

\vspace{1mm}
\emph{--- \badge{Tag Editor} badge:}
Stack Overflow users can include tags on questions (or answers) to concisely describe their content. 
In July 2010, 
Stack Overflow enabled the creation of tag wikis by the community, which aim to provide a description of all used tags. 
Shortly afterwards, it introduced a badge called \badge{Tag Editor}, 
awarded after a user edits a tag wiki for the first time, to encourage users to edit tag wikis.
To ensure the quality of the wiki tags, only users with at least a reputation level of $1{,}500$ could (initially) edit a tag 
wiki\footnote{\scriptsize In February 2011, Stack Overflow lowered the minimum reputation level to $100$ and thus the characteristics of the population
that could earn the badge changed. Therefore, in our analysis, we only consider data up to January 2010.}.
Finally, note that a user obtains a \emph{low} utility from editing a wiki tag---it requires some effort and she only receives the \emph{intangible} 
reward of helping the community. Moreover, the more uncommon a tag is, the least this intangible reward may be since she will help a smaller part
of the community.

\vspace{1mm}
\emph{--- \badge{Pro\-mot\-er} badge:}
When a Stack Overflow user does not receive a satisfactory answer to one of her questions, she can offer a \emph{bounty} to 
reward, in the form of reputation points, the user who would provide such a satisfactory answer\footnote{\scriptsize A user can also offer a bounty to a user after she has provided 
an answer, as a thank you gift, however, that usage is rarer.}.
In July 2010, Stack Overflow introduced a badge called \badge{Pro\-mot\-er}, awarded 
after a user offers a bounty for an answer to one of her questions for the first time, to encourage users to offer more bounties.
Only users with at least a reputation of $75$ points can offer a bounty.
In contrast with editing a wiki tag, a user obtains a \emph{high} utility from offering a bounty---it requires little effort and she may receive an answer to a question
she is personally interested in, however, it entails a cost in terms of the reputation she transfers to the user providing the answer.

\vspace{1mm}
\emph{--- \badge{In\-ves\-tor} badge:}
Stack Overflow users can also offer bounties to receive a satisfactory answer to a question that has been asked by another user. In July 2010, Stack 
Overflow introduced a first-time badge called \badge{In\-ves\-tor} to encourage users 
to offer more bounties for answers to other users'{} questions.
Similarly as in the \badge{Pro\-mot\-er} badge, only users with at least a reputation of $75$ points can offer a bounty for an answer to a question asked by another user. 
However, in this case, a user may obtain a \emph{lower} utility from offering a bounty for an answer to another user'{}s question than her own---on the one hand, she 
may be less interested in an answer since she did not originally ask the question and, on the other hand, the question may have already a (relatively) satisfactory answer 
when she found it.

\vspace{1mm} \noindent Table~\ref{tab:data} provides general statistics on the number of eligible users and user actions in our dataset.

\begin{table}[!t]
\caption{General statistics on the number of users and actions in our dataset. The time when a badge is introduced is denoted by $\tau$, the time when a user becomes eligible
to perform an action is denoted by $s_u$ and the time when a user performs the action is denoted by $t_u$. \vspace{-3mm}} 
\centering
\small
\scalebox{1.0}{
\begin{tabular}{ l | c c c } 
 & Tag Editor$^{*}$ & Investor & Promoter\\
 \hline  
Required reputation & $1{,}500$ & $75$ & $75$ \\
Eligible users, $s_u < \tau$ & $6{,}396$ & $46{,}148$ & $46{,}148$ \\
Actions, $t_u < \tau$ & $93$ & $4{,}984$ & $205$ \\
Eligible users, $s_u \geq \tau$ & $2{,}095$ & $297{,}481$ & $297{,}481$ \\
Actions, $t_u \geq \tau$ & $471$ & $30{,}396$ & $7{,}830$ \\
 \hline  
\end{tabular}
}
{\footnotesize
\begin{tabular}{ l}
\vspace{-2mm}
\\$^{*}$For \badge{Tag Editor} badge we present statistic only until 2011-02-09 \\ when the required reputation changed. \\
\end{tabular}
}
\label{tab:data}
\vspace{-3mm}
\end{table}

\section{Testing the effectiveness of badges}
\label{sec:framework}
In this section, we first formalize the problem setting, which includes at its core a natural experiment. Then, we introduce our causal inference framework for first-time
badges, discussing the assumptions it requires and elaborating on its individual components.
Finally, we evaluate its effectiveness using a variety of synthetic experiments and conclude with a discussion of its limitations.

\xhdr{Problem setting} 
Given an action of interest $a$, we record the behavior of each user during an observation window $[0, T]$ as a tuple
\begin{equation}
e~~\defeq~~(\explainup{s_u}{\text{start time}}, \quad \explain{t_u}{\text{action time}}, \quad \explainup{{\color{gray} {v_u}}}{\text{utility}}~), 
\end{equation}
which means that user $u$ becomes eligible\footnote{\scriptsize Stack Overflow often requires a minimum reputation level for a user to be able to perform 
an action. For example, only users with at least a reputation level of $75$ can offer a bounty.} to perform the action at time $s_u$, she performs the action at 
time $t_u$, with $t_u \geq s_u$, and obtains a utility $v_u$, which is often {\color{gray} intangible}. 
If a user does not perform the action during the observation window $[0, T]$, we set the action time to $t_u = \infty$, 
however, this does not imply she will never perform the action.
Moreover, we assume a first-time badge $b$ is introduced at time $\tau \in (0, T]$ to incentivize users to take action $a$. That means, after time $\tau$, a user 
receives badge $b$ the first time she takes action $a$, potentially \emph{increasing} its corresponding utility $v_u$.
Here, the \emph{delayed} introduction of badge $b$ at time $\tau$ can be thought of as a natural experiment~\cite{meyer1995natural, rosenzweig2000natural}, 
where the treatment of earning a badge is assigned \emph{as if random} among users who are eligible to perform an action.
Note that users who performed action $a$ before the badge introduction time $\tau$ receive the badge automatically at $\tau$, however, this does not influence our analysis 
since our focus is on the potential effect that first-time badges may have on the first time users take the action, and not on subsequent times.

Given the above setup, our goal is then to assess to which extent the introduction of the badge \emph{changes} users'{} behavior, as measured by the time users take
to perform the action for the first time, \ie, $t_u - s_u$.
Next, we present a high-level overview of our causal inference framework, highlighting the assumption it requires, and then elaborate on its individual components.

\xhdr{Our causal inference framework}
Given an action of interest $a$, its corresponding first-time badge $b$ with introduction time $\tau$, the behavior of $n$ users 
with respect to $a$, \ie, $\Dcal_a = \{ (s_u, t_u, v_u) \}_{u \in [n]}$, 
one could just consider the set of users who performed the action for the first time before the badge introduction, \ie, $t_u < \tau$, as 
control group, and the set of users who became eligible to perform the action after the badge introduction, \ie, $\tau < s_u$, as 
treatment group. 
However, proceeding that way would have several limitations. 
First, a potential difference among the treatment and control groups could be due to random fluctuations in users'{} behavior over time 
due to many complex, confounding factors.
Second,  one would be unable to consider users who became eligible before the badge introduction but performed the action for the first time after 
the badge introduction. These users are the ones who were actually exposed to the badge introduction.

Our causal inference framework overcomes the above limitations by redefining the treatment and control groups as follows.
We define the treatment group as the set of users who became eligible to perform the action at a time 
$s_u \in [\tau - w/2, \tau + w/2]$, where $w \geq 0$ is a given parameter. 
We define several control groups, each of them associated to a \emph{virtual} badge $b_i$ introduced at time 
$\tau_i \in [w/2, \tau-w] \cup [\tau+w, T-w/2]$, picked uniformly at random.
Under these definitions, the treatment group represents the \emph{true world} where the badge is introduced and the control groups 
re\-pre\-sent different \emph{counterfactual worlds} where a virtual badge, which models random fluctuation in users'{} behavior, is introduced.
As long as the allocation of users across the treatment and control groups resemble random assignment, 
if the strength of the change induced by the badge in the true world (as measured by survival-based hypothesis testing) is larger than the strength 
of the changes induced by the virtual badges in a variety of counterfactual worlds, we can conclude that the change in the true 
world is not due to random fluctuation in users'{} behavior with higher confidence. 

Next, we elaborate on two survival-based hypothesis testing procedures of increasing statistical power, which we use to measure the strength of the change induced by
a (true or virtual) badge, and then describe how to formally compare the strength of the changes induced by the true and the virtual badges by means 
of a novel bootstrap difference in differences method.

\vspace{1mm}
\noindent\hspace{0mm} \emph{--- Basic survival-based hypothesis testing: }
Given an action of interest $a$, we model the time $t_u$ when a user $u$ takes action $a$ using a survival process~\cite{Aalen2008}. Following the literature on temporal point processes, we represent such
survival process as a binary counting process $N_u(t) \in \{0, 1\}$, which becomes one when the user performs the action for the first time. Then, we characterize this counting process using its corres\-pon\-ding 
intensity $\lambda_u(t)$, \ie, $\EE[dN_u(t)] = \lambda_u(t) dt$, which we define as follows:
\begin{equation}
\lambda_u(t) =
\begin{cases}
  0 & \text{if}\enskip t < s_u \\
  {\lambda_0} & \text{if}\enskip s_u \leq t < \tau\\
  {\lambda_1} & \text{otherwise}
\end{cases}
\label{eq:survival-model}
\end{equation}
where $s_u$ is the time when user $u$ becomes eligible to perform action $a$, $\tau$ is the time when the first-time badge is introduced, and 
$\lambda_0$ and $\lambda_1$ are parameters shared across all users, which depend on the (intangible) utility users obtain from taking the action
for the first time before and after the badge introduction, respectively.

Under this model, the null hypothesis $\Hcal_0$, \ie, the badge did not have an effect, corresponds to $\lambda_0 = \lambda_1 \geq 0$ and the alternative hypothesis $\Hcal_1$ is $\lambda_0 \neq \lambda_1$ 
with $\lambda_0 \geq 0$ and $\lambda_1 \geq 0$.
Moreover, given the behavior of $n$ users, the maximum likelihood estimators of the model parameters, $\hat{\lambda}_0$ and $\hat{\lambda}_1$, can be computed analytically. In particular, under the null 
hypothesis, they are readily given by:
\begin{equation}
\hat{\lambda}_0 = \hat{\lambda}_1 = \frac{\sum_{u \in [n]} \II(t_u \leq T)}{\sum_{u \in [n]} ( \min(t_u, T) - s_u)}, \label{eq:lambda-basic-mle-null}
\end{equation}
and under the alternative hypothesis, they are given by:
\begin{align}
\hat{\lambda}_0 &= \frac{\sum_{u \in [n]} \II(t_u \leq \tau)}{\sum_{u \in [n]} ( \min(t_u, \tau) - s_u ) \II(s_u < \tau)} \nonumber \\
\hat{\lambda}_1 &= \frac{\sum_{u \in [n]} \II(\tau < t_u \leq T)}{\sum_{u \in [n]} ( \min(t_u, T) - \tau ) \II(t_u > \tau)}, \label{eq:lambda-basic-mle-alternative}
\end{align}
where $\II(\cdot)$ is the indicator function and all the sums are over eligible users. Then, we can use a test statistic such as standard log-likelihood ratio ($\textit{LLR}$)~\cite{hogg1995introduction}
to measure the strength of the change induced by the badge, \ie, 
\begin{align*}
LLR &= \sum_{u \in [n] : t_u < T} \log f(s_u, t_u ; \hat{\lambda}_0, \hat{\lambda}_1) - \log f(s_u, t_u ; \hat{\lambda}_0)  \\  
&+ \sum_{u \in [n] : t_u = \infty} \log S(s_u, T ; \hat{\lambda}_0, \hat{\lambda}_1)  - \log S(s_u, T ; \hat{\lambda}_0)
\end{align*}
where $f(s_u, t_u ; \hat{\lambda}_0)$ and $f(s_u, t_u; \hat{\lambda}_0, \hat{\lambda}_1)$ are the likelihoods of the action times under
the null and alternative model, respectively, and $S(s_u, T ; \hat{\lambda}_0, \hat{\lambda}_1)$ and $S(s_u, T ; \hat{\lambda}_0)$ are
the corresponding survival functions.
Here, the likelihoods and survivals can be computed using the intensity $\lambda_u(t)$ defined by Eq.~\ref{eq:survival-model}, with 
$\lambda_0 = \lambda_1$ for the null model and $\lambda_0 \neq \lambda_1$ for the alternative model, as noted elsewhere~\cite{Aalen2008}.

\vspace{1mm}
\noindent\hspace{0mm} \emph{--- Robust survival-based hypothesis testing: }
The basic sur\-vi\-val-based hypothesis testing procedure described above assumes the model parameters are shared 
across all users and, by doing so, it ignores the utility heterogeneity across users.
Here, 
we account for the utility heterogeneity by considering different latent parameters per user, but sampled from the same distributions, \ie,
\begin{align}
\lambda_u(t) &=
\begin{cases}
  0 & \text{if}\enskip t < s_u \\
  {\lambda_0(u)} & \text{if}\enskip s_u \leq t < \tau\\
  {\lambda_1(u)} & \text{otherwise}
\end{cases} \nonumber \\
\lambda_0(u) &\sim \mbox{Gamma}(k_0, r) \nonumber \\
\lambda_1(u) &\sim \mbox{Gamma}(k_1, r)
\label{eq:robust-survival-model}
\end{align}
where $s_u$ is the time when user $u$ becomes eligible to perform action $a$, $\tau$ is the time when the first-time badge is introduced, 
$k_0, k_1$ are shape parameters and $r$ is a rate parameter. Here, note that $\EE[\lambda_0] = k_0 / r$ and $\EE[\lambda_1] = k_1 / r$.

Then, we define the null and alternative hypothesis in terms of the shape parameters, \ie,
\begin{align*}
\Hcal_0 & \,:\, k_0 = k_1 \geq 0 \\
\Hcal_1 & \,:\, k_0 \neq k_1, k_0 \geq 0, k_1 \geq 0
\end{align*}
Moreover, given the behavior of $n$ users, we can estimate the shape parameters using maximum likelihood estimation, integrating out 
the latent parameters $\lambda_0(u)$ and $\lambda_1(u)$, and estimate the rate parameter by cross vali\-da\-tion. 
More specifically, under the null hypothesis, the shape parameters are given by:
\begin{equation}
\hat{k}_0 = \hat{k}_1 = -\frac{\sum_{u \in [n]} \II(t_u \leq T)}{ \sum_{u \in [n]} \log \left( \frac{r}{r+\min(t_u, T) - s_u} \right) } \label{eq:lambda-robust-mle-null}
\end{equation}
and under the alternative hypothesis, they are given by:
\begin{align}
\hat{k}_0 &= -\frac{\sum_{u \in [n]} \II(t_u \leq \tau)}{ \sum_{u \in [n]} \log \left(\frac{r}{r+\min(t_u, \tau) - s_u}\right) \II(s_u \leq \tau) } \nonumber \\
\hat{k}_1 &= - \frac{\sum_{u \in [n]} \II(\tau < t_u \leq T)}{\sum_{u \in [n]}  \log\left(\frac{r}{r+\min(t_u, T) - \tau}\right) \II(t_u > \tau) } \label{eq:lambda-robust-mle-alternative}
\end{align}

Similarly as in the basic survival model, we can then use a test statistic such as standard log-likelihood ratio ($\textit{LLR}$) to measure
the strength of the change induced by the badge.

\begin{figure}[t!]
\centering
\includegraphics[width=0.315\textwidth]{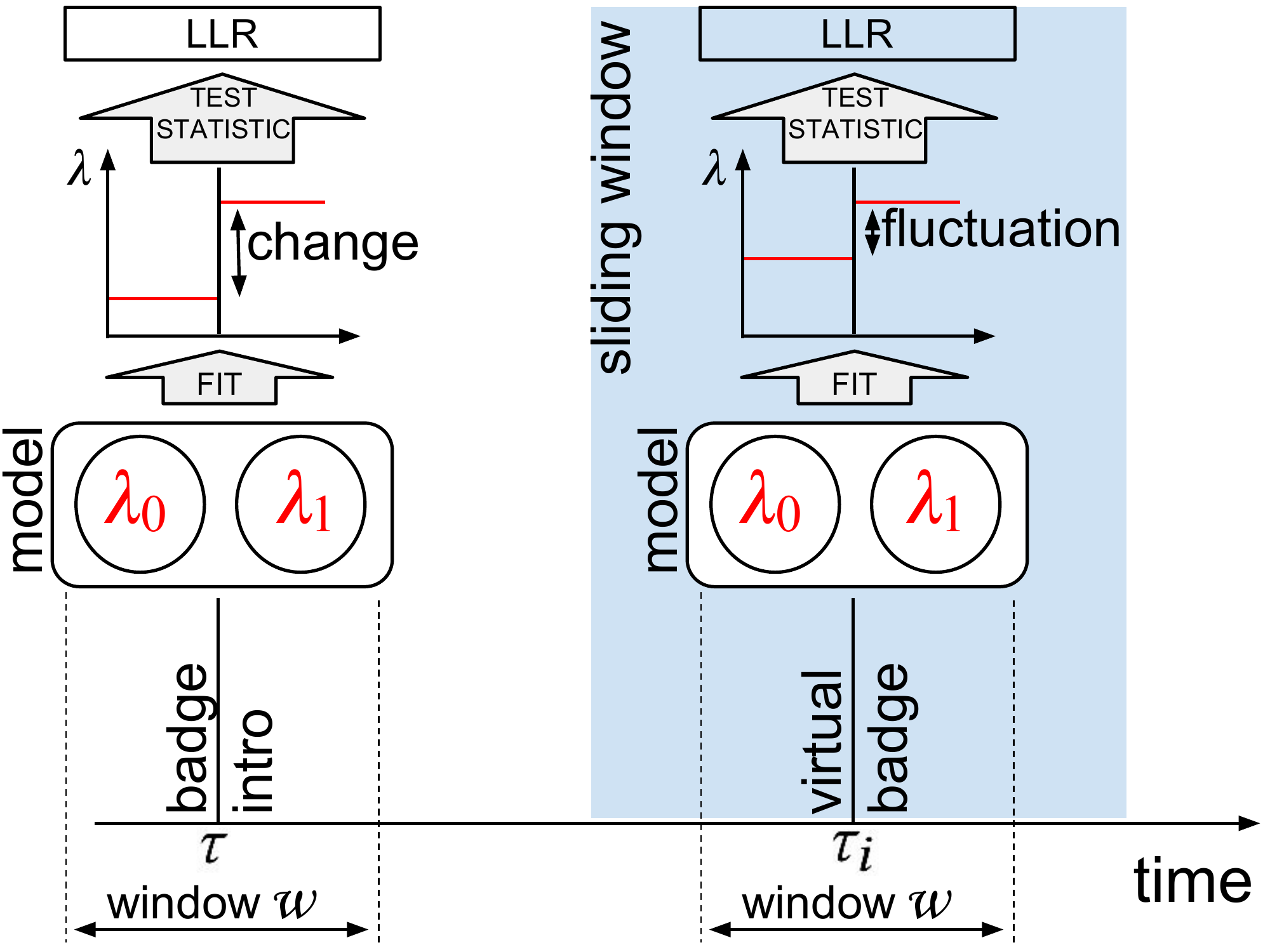}
\caption{Our bootstrap difference-in-difference method. 
The treatment group (left) consists of users whose start time $s_u$ lies in a window of size $w$ around the time $\tau$ the badge is introduced. 
Each control group $i$ (right) consists of users whose start time $s_u$ lies in a window of size $w$ around the time $\tau_i$ a virtual badge is introduced. 
The method runs hypothesis testing on both the treatment group and all control groups and then compares the test statistic (\eg, LLR) of the treatment 
group with the empirical distribution of the test statistic of the control groups.}
\label{fig:framework}
\vspace{-3mm}
\end{figure}

\vspace{1mm}
\noindent\hspace{0mm} \emph{--- Bootstrap difference-in-differences method: } 
Given an action of interest $a$, its corresponding first-time badge $b$ with introduction time $\tau$, the behavior of $n$ users with respect 
to $a$, \ie, $\Dcal_a = \{ (s_u, t_u, v_u) \}_{u \in [n]}$, and a model-based hypothesis testing procedure as the ones described above,
we assess the significance of the change induced by the true badge using the following bootstrap difference-in-difference method:
\vspace{1mm}
\begin{itemize}[noitemsep,nolistsep,leftmargin=0.7cm]
\item[I.] We select all users whose start time $s_u \in [\tau-w/2, \tau+w/2]$, where $w \geq 0$ is a given parameter, as treatment group. Then, we compute 
the maximum likelihood estimators of the parameters of the model of choice using Eqs.~\ref{eq:lambda-basic-mle-null}-\ref{eq:lambda-basic-mle-alternative} 
(basic) or Eqs.~\ref{eq:lambda-robust-mle-null}-\ref{eq:lambda-robust-mle-alternative} (robust) on that subset of users and use these estimated parameters 
to compute the log-likelihood ratio, $LLR$.

\item[II.] We introduce a set of virtual badges $\Vcal$ at a times $\tau_i \in [w/2, \tau-w] \cup [\tau+w, T-w/2]$, picked uniformly at random (in practice, one can use a sliding 
window), where $w \geq 0$ is the same given parameter as in the first step. Then, for each virtual badge $i \in \Vcal$, we select users whose start time $s_u \in [\tau_i - w/2, \tau_i + w/2]$ as control group $i$.
Finally, we compute the maximum likelihood estimators of the parameters of the model of choice, as in the previous step, on each of these subsets of users and use these estimated parameters 
to compute the log-likelihood ratio for each control group, $LLR_{\tau_i}$. 

\item[III.] We measure the strength of the change induced by the badge by means of the probability that the LLR of the control groups, for which the null hypothesis 
holds by design since virtual badges do not exist, is larger than the LLR of the treatment group, $p := F_{LLR}(LLR_{\tau})$. 
\end{itemize}
\vspace{1mm}
The above bootstrap difference-in-differences method, which we also illustrate in Figure~\ref{fig:framework}, equips us with a robust empirical
estimate of the distribution of the LLR under the null hypothesis $F_{LLR}(LLR)$ and a $p$-value $p = F_{LLR}(LLR_{\tau})$, which accounts 
for the temporal fluctuations in users'{} behavior and allows us to reject the null hypothesis with higher confidence.
Finally, as discussed previously, the main assumption needed for the above method to be valid is that the allocation of users across the treatment 
and control groups resemble random assignment. Table~\ref{tab:smd} shows this assumption is satisfied in our dataset, as we will discuss later.
\begin{figure}[t]
\centering
\captionsetup[subfigure]{labelformat=empty}
\captionsetup[subfigure]{justification=centering}

\subfloat[Average $p$-value]{
\hspace*{-2mm}
\includegraphics[width=0.315\textwidth]{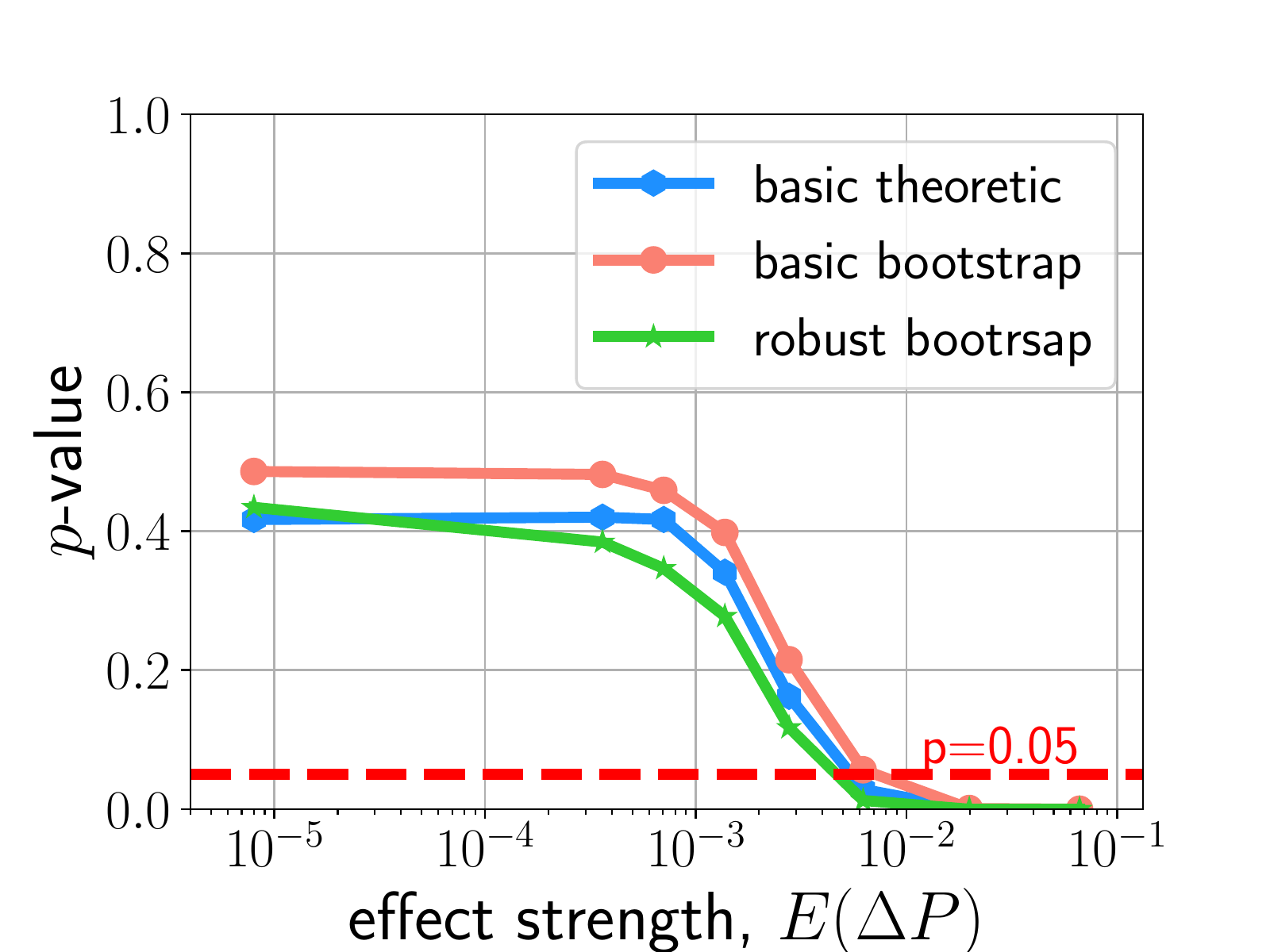}
}
\subfloat[Rejection probability]{
\hspace*{-5mm}
\includegraphics[width=0.315\textwidth]{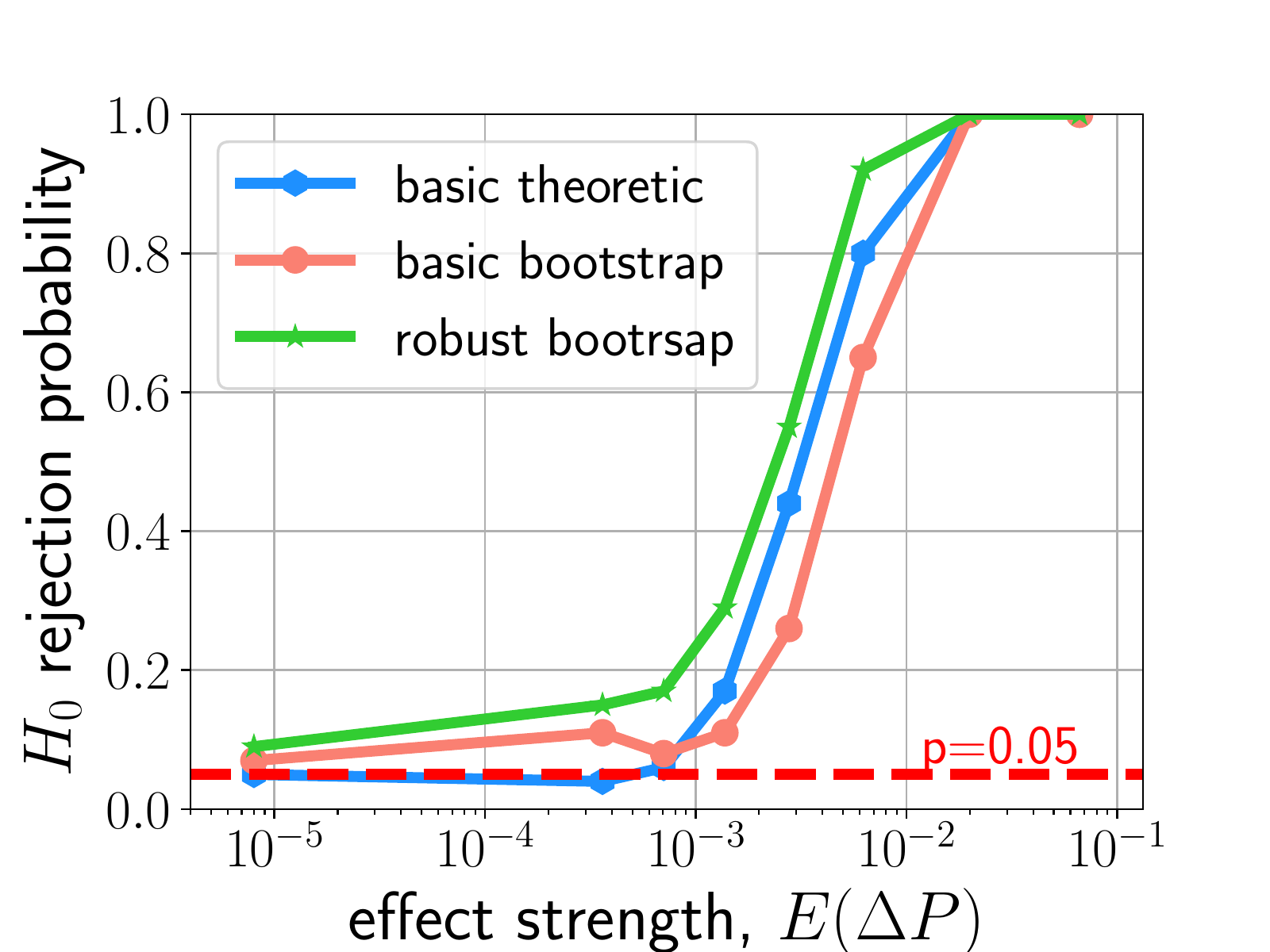}
}
\caption{Performance of our causal inference framework on synthetic data. The left panel shows the average $p$-value against expected effect strength $\EE[\Delta P]$, where lower (higher) is better 
for $\EE[\Delta P] > 0$ ($\EE[\Delta P] \approx 0$). The right panel  informs about tests' statistical power by
showing the rejection probability of the null hypothesis $\Hcal_0$ at $p = 0.05$ against effect strength $\EE[\Delta P]$, where higher (lower) is better for $\EE[\Delta P] > 0$ ($\EE[\Delta P] \approx 0$).}
\label{fig:synthetic}
\vspace{-3mm}
\end{figure}

\xhdr{Framework evaluation}
In this section, we compare the effectiveness of the basic survival model with the theo\-re\-ti\-cal distribution of the LLR under the null hypothesis ($\chi_1^2$, given by Wilks' theorem~\cite{wilks1938large}; 
``basic theoretical'') and the basic and robust survival models with the empirical distribution ($F_{LLR}$) of the LLR under the null hypothesis, as estimated by the proposed difference-in-differences 
bootstrap method (``basic bootstrap'' and ``robust bootstrap'', respectively).
More specifically, we proceed as fo\-llows.

First, we simulate the behavior of $n = 10{,}000$ users during a time interval $[0, T]$, where $T = 360$. For each user, we draw her starting times $s_u \sim U[0, T]$, and her action time $t$ from an intensity $\lambda_u(t) (1+at)$, where $\lambda_u(t)$ is given by Eq.~\ref{eq:robust-survival-model}
and $a = 0.001$. 
Note that the term $(1+at)$ imposes a global linear trend, which is often observed in real data\footnote{\scriptsize We obtain quantitatively similar results in the absence
of a linear trend, \ie, $a = 0$.}.
Moreover, in Eq.~\ref{eq:robust-survival-model}, we set the badge introduction time to $\tau = T/2$, the rate parameter to $r = 10$, and consider different 
badge strength values expressed as the expected increase in the probability that users perform the action in less than $t = 10$ after $s_u$, which we denote
as $\EE[\Delta P]$.
For each configuration, we run $100$ independent simulations.

Then, we run the above methods (``basic theoretical'', ``basic bootstrap'', and ``robust bootstrap'') on data from each of the independent simulations 
and measure their effectiveness in terms of two metrics: average $p$-value and rejection probability of the null hypothesis $\Hcal_0$ at $p=0.05$.
Figure~\ref{fig:synthetic} summarizes the results,
which show that the robust bootstrap has a superior performance: it is more likely to reject $\Hcal_0$ when a badge is introduced (\ie, $\EE[\Delta P] > 0$) while it is equally 
likely not to reject $\Hcal_0$ when a badge is not introduced (\ie, $\EE[\Delta P] \approx 0$).

\xhdr{Remarks}
We acknowledge that our framework has several limitations. 
In its current form, the framework is only applicable to first-time badges. In principle, it may be possible to augment our framework to other types of badges. 
For example, in threshold badges, one could replace the binary counting processes by non binary ones representing the number of actions $N(t)$ (\ie, progress) and choose a functional
form for the intensities depending on the numbers of actions, \ie, $\lambda_1(N(t))$, as suggested in previous work~\cite{anderson2013steering}. 
However, one would need to address additional challenges, \eg, accurate intensity models, and it is out of the scope of this work.
Moreover, while our bootstrap difference-in-difference method does control for temporal fluctuations in the users'{} behavior, it cannot rule out the possibility that the change in users'{} behavior 
in the treatment group, if deemed significant, may be due to a hidden confounding factor rather than the badge introduction.

\section{Do badges work?}
\label{sec:results}

Before we apply our causal inference framework to the three first-time badges described in Section~\ref{sec:data}, we first check the random assignment
assumption between the treatment and control groups, which is necessary for our framework to provide sound conclusions.
If the assignment is random, we would expect the groups (respective subpopulations) to be indistinguishable based on any additional covariates. Two groups are called
indistinguishable (or balanced) if these additional covariates are within a standardized mean difference (SMD)\footnote{\scriptsize The standardized mean
difference (SMD) is defined as the difference in means of the treatment and control group divided by the pooled standard deviation of both groups.} 
of 0.25 standard deviations~\cite{stuart2010matching}.
In particular, Table~\ref{tab:smd} shows that treatment and control groups are indistinguishable in our dataset.
Here, we would like to acknowledge that conclusions drawn from standardized mean differences (SMD) when eva\-lua\-ting variables with skewed distributions, as ours, 
should be taken with caution.

\begin{figure*}[t]
\centering             
\captionsetup[subfigure]{labelformat=empty}
\captionsetup[subfigure]{justification=centering}

\rotatebox{90}{\scriptsize \hspace*{1.0cm} \textbf{\badge{Tag Editor}}}
\subfloat[]{
\includegraphics[width=0.27\textwidth]{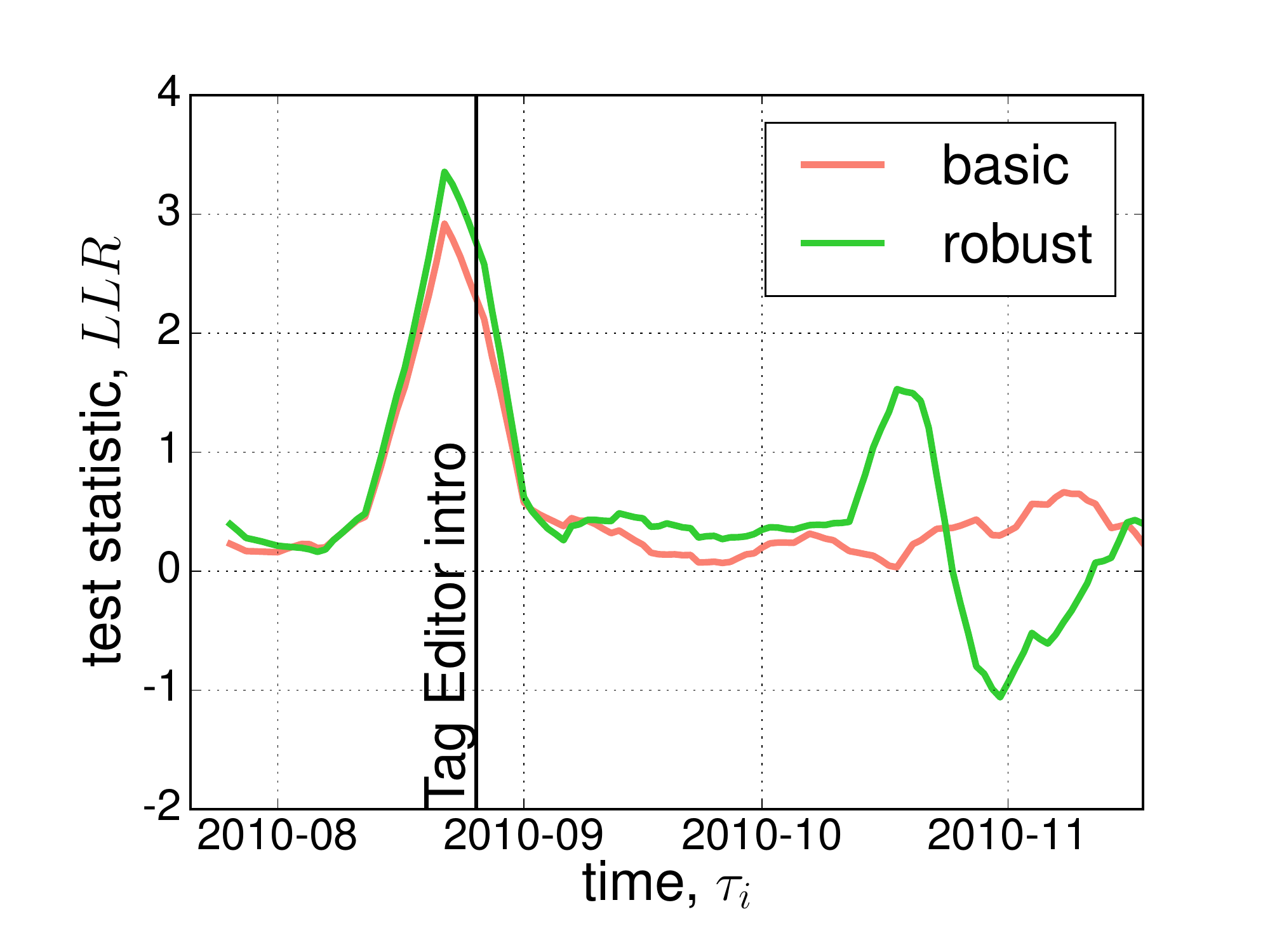} 
}\hspace*{0.3cm}
\subfloat[]{
\includegraphics[width=0.27\textwidth]{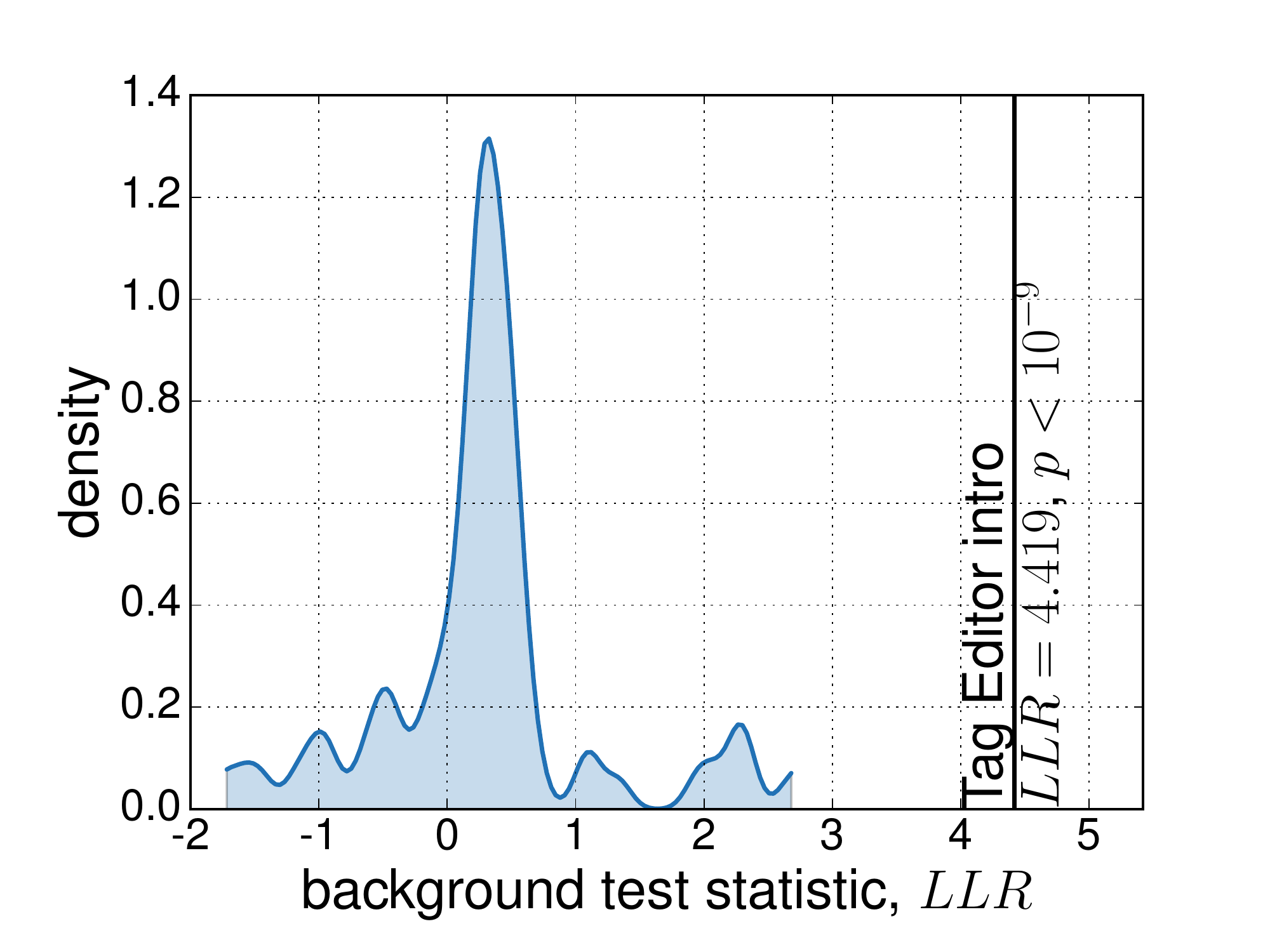}
}\hspace*{0.3cm}
\subfloat[]{
\includegraphics[width=0.27\textwidth]{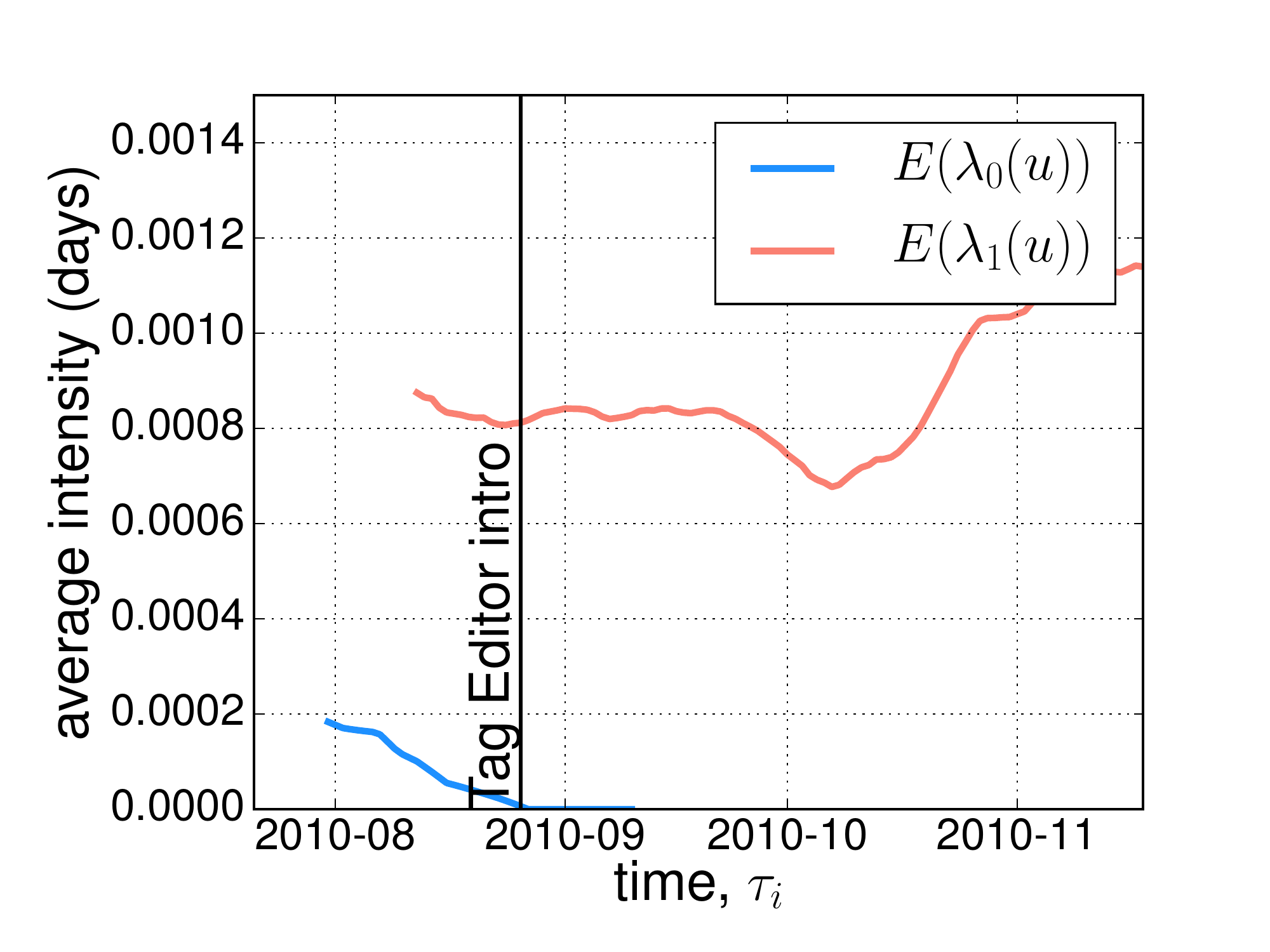}
}

\vspace{-0.9cm}

\rotatebox{90}{\scriptsize \hspace*{1.35cm} \textbf{\badge{Pro\-mot\-er}}}
\subfloat[]{
\includegraphics[width=0.27\textwidth]{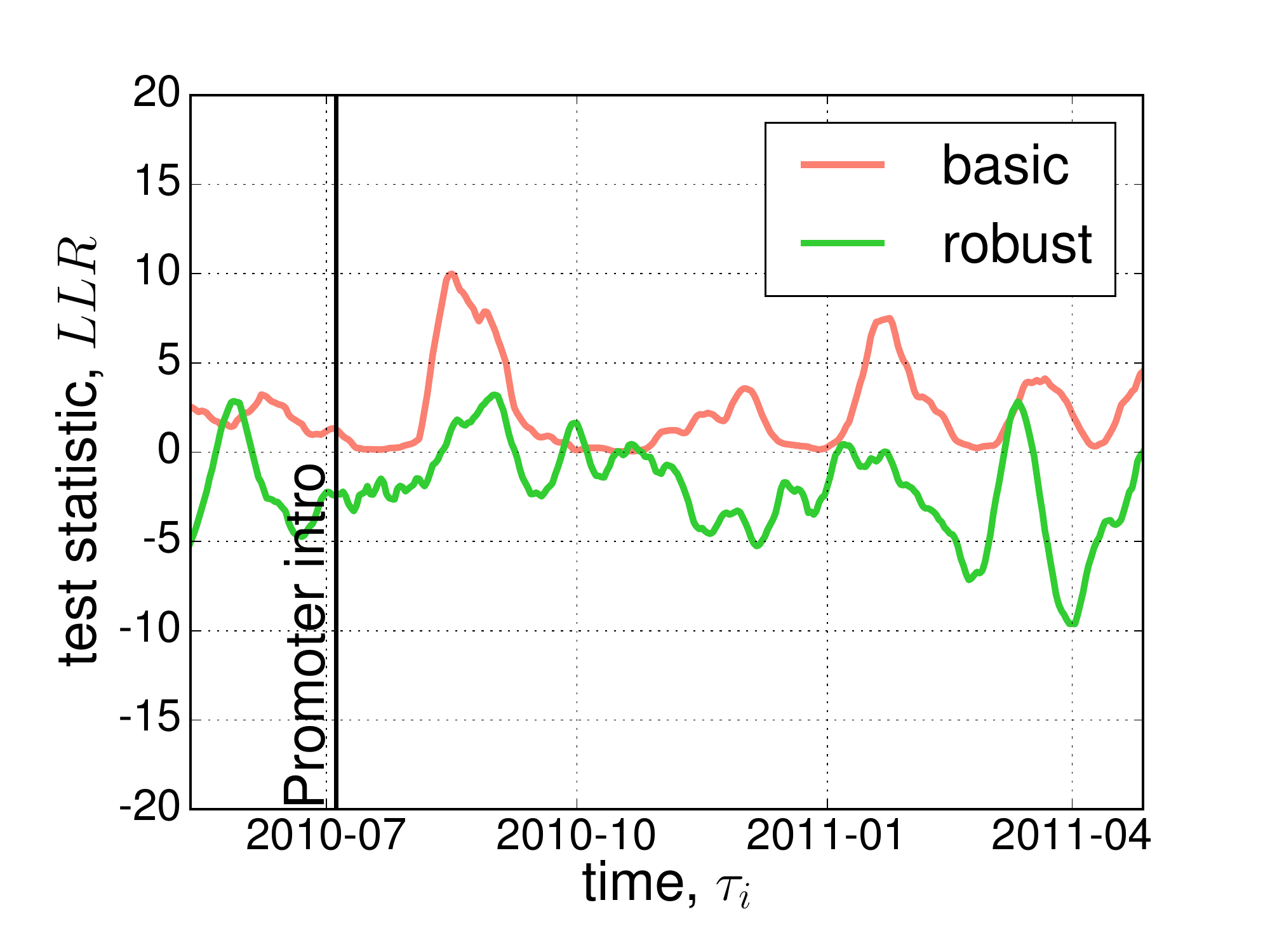} 
}\hspace*{0.3cm}
\subfloat[]{
\includegraphics[width=0.27\textwidth]{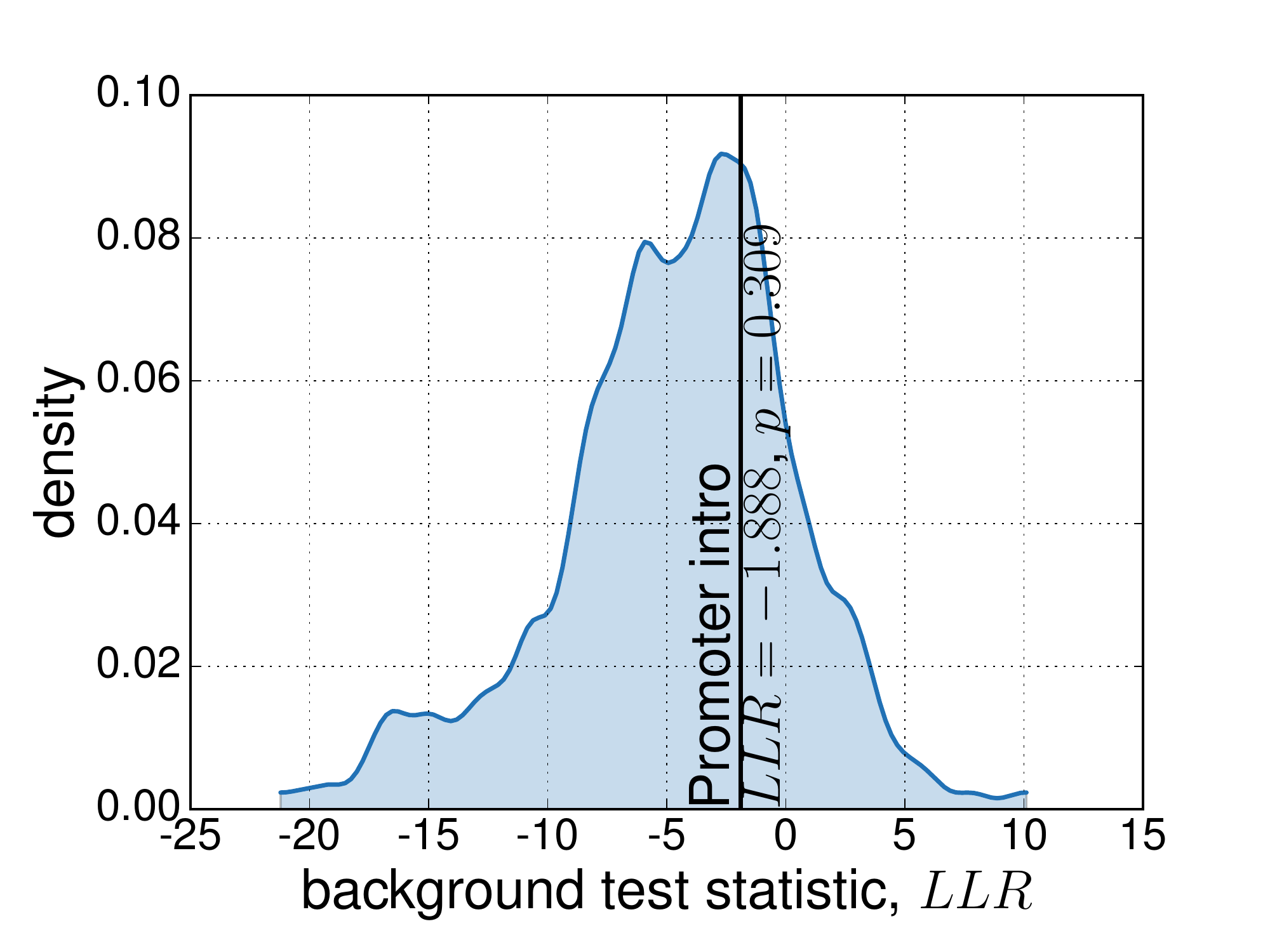}
}\hspace*{0.3cm}
\subfloat[]{
\includegraphics[width=0.27\textwidth]{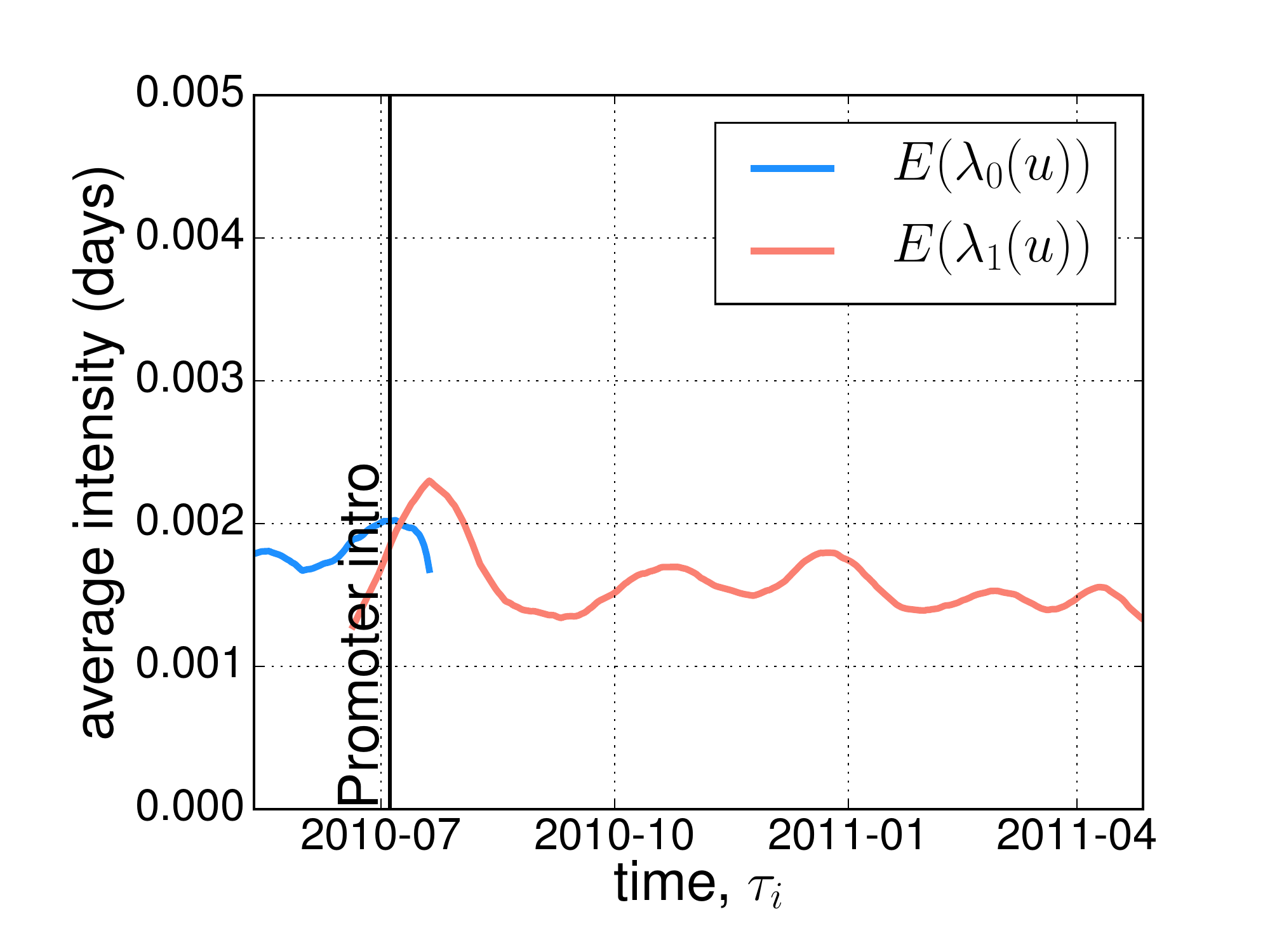}
}

\vspace{-0.9cm}

\rotatebox{90}{\scriptsize \hspace*{1.35cm} \textbf{\badge{In\-ves\-tor}}}
\subfloat[$LLR$ over time (basic~and~robust)]{
\includegraphics[width=0.27\textwidth]{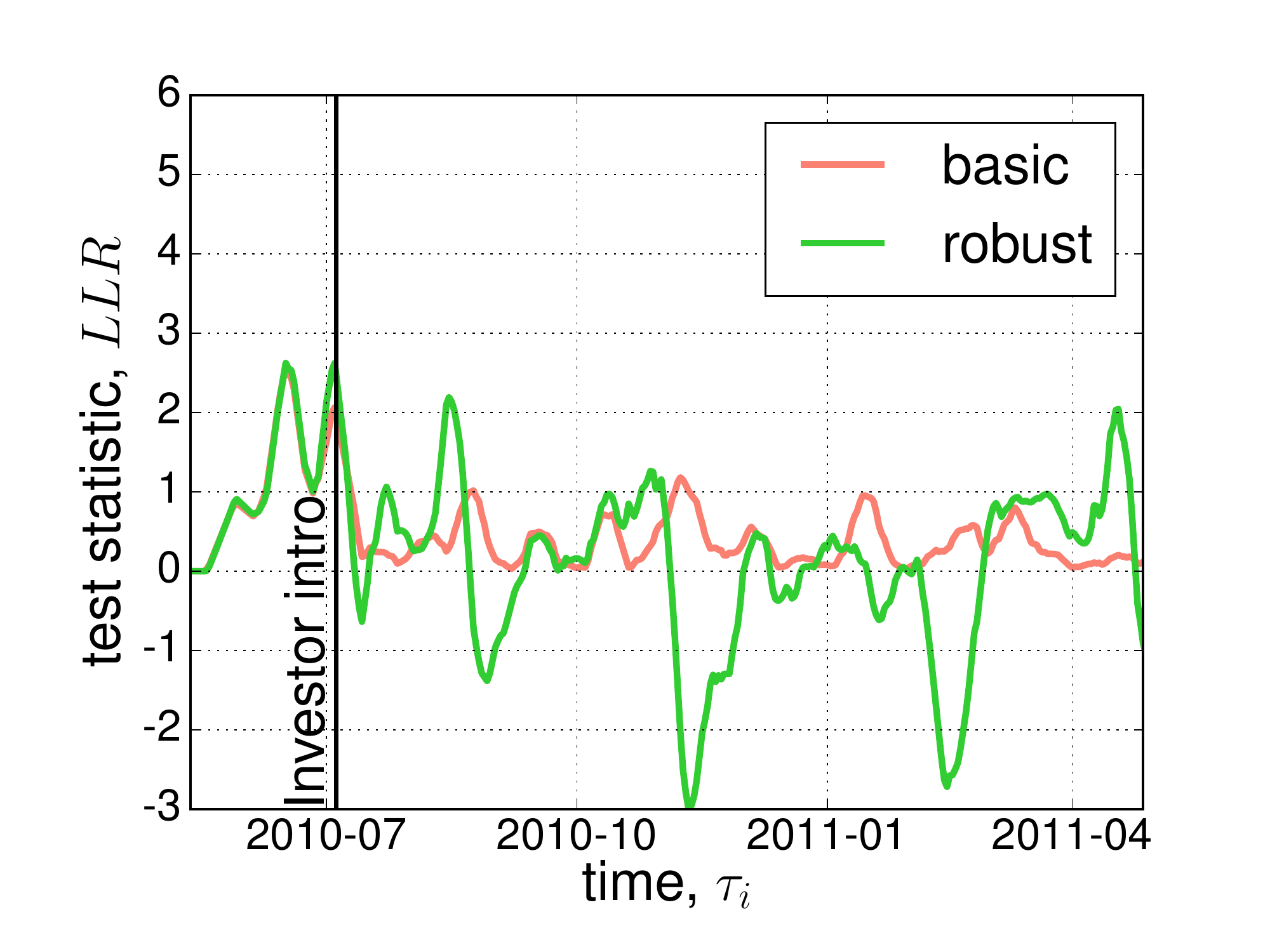}
}\hspace*{0.3cm}
\subfloat[$LLR$ distribution under $\Hcal_0$ (robust)]{
\includegraphics[width=0.27\textwidth]{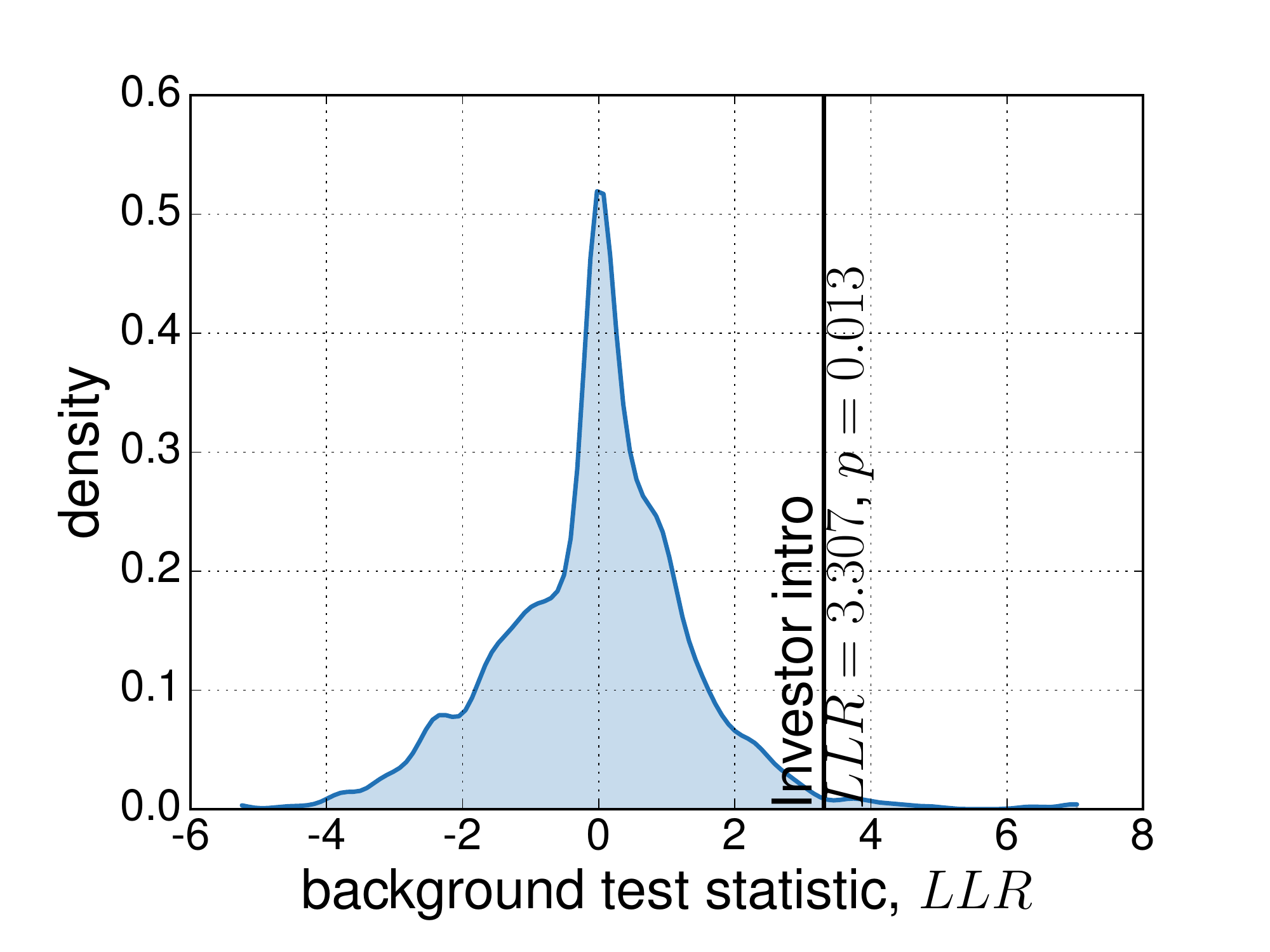}
}\hspace*{0.3cm}
\subfloat[Action intensity over time (robust)]{
\includegraphics[width=0.27\textwidth]{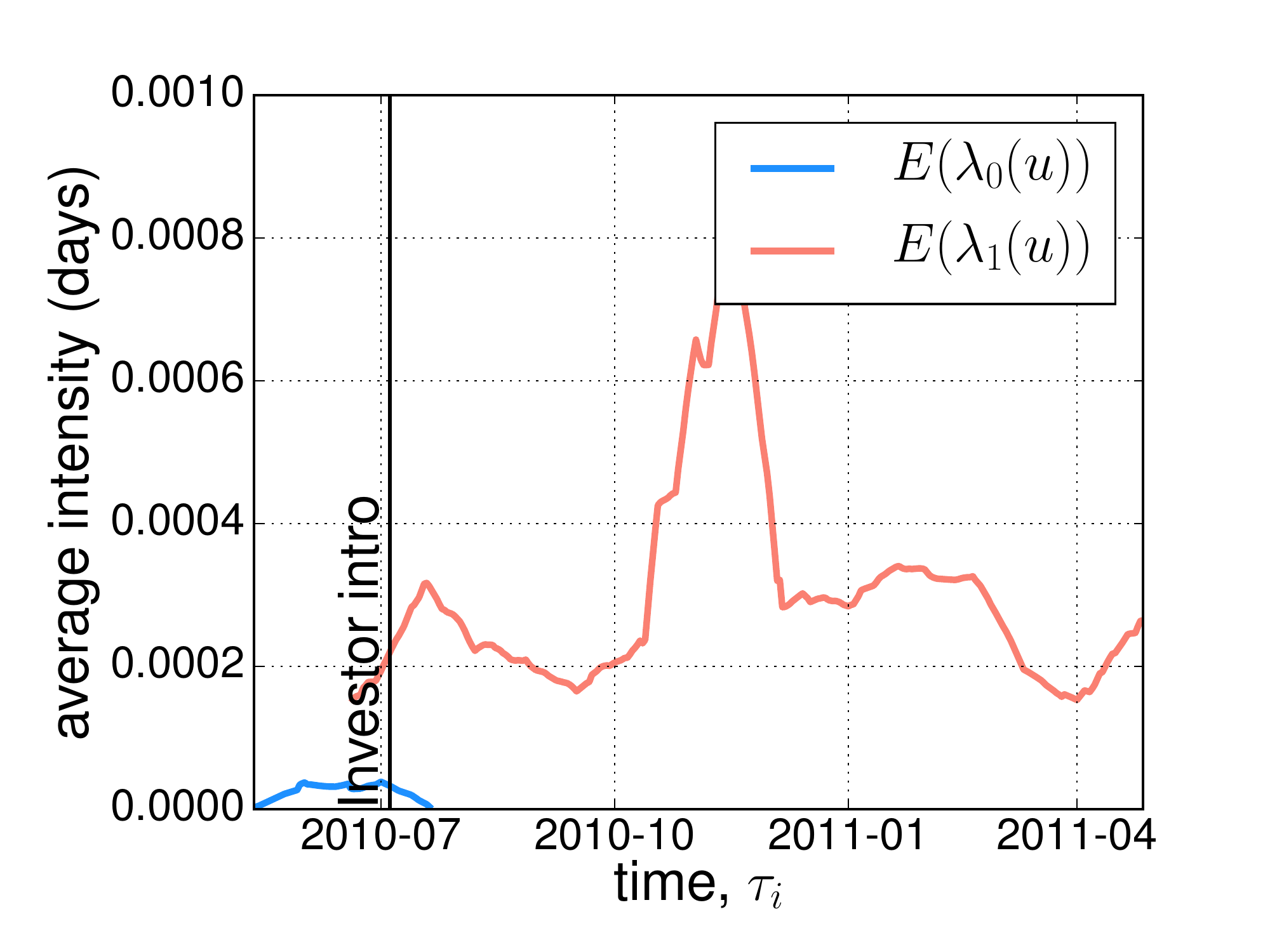}
}
\caption{Causal effect of three badges (\badge{Tag Editor}, \badge{Pro\-mot\-er} and \badge{In\-ves\-tor}) on users'{} behavior.
Panels in the left show the test statistic (log-likelihood ratio; $LLR$) over time for the basic and robust survival models, \ie,
$LLR_{\tau_i}$ and $LLR_{\tau}$ against $\tau_i$ and $\tau$.
Panels in the middle show the empirical distribution of the test statistic ($LLR$) under the null hypothesis $\Hcal_0$ and
$p$-value for the robust survival model, \ie, $F_{LLR}(LLR)$ and $p = F_{LLR}(LLR_\tau)$.
Panels in the right show the average intensities for first-time action under robust survival model, \ie, $\hat{k}_0$ before $\tau$
and $\hat{k}_1$ after $\tau$, using a sliding window of length $w = 60$ days.
The results suggest that the \badge{Tag editor} and \badge{In\-ves\-tor} badges were \emph{successful} and the \badge{Pro\-mot\-er}
badge was \emph{unsuccessful}.
}
\label{fig:user_results}
\vspace{-2mm}
\end{figure*}

Once we have validated the random assignment assumption, we apply our causal framework to the three first-time badges. Figure~\ref{fig:user_results} 
summarizes the results by means of: 
\begin{itemize}[noitemsep,nolistsep,leftmargin=0.75cm]
\item[(i)] Test statistic over time for the basic and robust survival models, \ie, $LLR_{\tau_i}$ and $LLR_{\tau}$ against $\tau_i$ and $\tau$.
\item[(ii)] Empirical distribution of the test statistic under $\Hcal_0$ and $p$-value for the robust survival model, \ie, $F_{LLR}(LLR)$  and $F_{LLR}(LLR_{\tau})$.
\item[(iii)] Average intensities for first-time action under robust survival model, \ie, $\hat{k}_0$ before $\tau$ and $\hat{k}_1$ after $\tau$, using a sliding window of length 
$w = 60$ days.
\end{itemize}

\noindent Overall, the results suggest that the \badge{Tag editor} and \badge{In\-ves\-tor} badges were \emph{successful}---they had a significant 
causal effect on users'{} behavior ($p = 0.004$ and $p = 0.017$, respectively). In contrast, the \badge{Pro\-mot\-er} badge was unsuccessful---it did 
not have a significant causal effect ($p = 0.309$). Moreover, a detailed analysis also reveals several interesting insights. 

First, the actions rewarded by the two successful badges were rare by the time the badges got introduced.
For example, in the case of the \badge{Tag editor} badge, only $100$ tag wiki edits had been performed, however, there were $\sim$$6{,}500$ users 
who were eligible to perform edits.
In the case of the \badge{In\-ves\-tor} badge, only $40$ bounties had been offered for an answer to other users'{} questions.
In contrast, the action rewarded by the \badge{Pro\-mot\-er} badge---offering a bounty for an answer to the users'{} own questions---was much more common by 
the time the badge got introduced. As a consequence, the average intensity for the \badge{Pro\-mot\-er} badge was an order of magnitude higher than the intensities corresponding to the 
\badge{Tag editor} or the \badge{In\-ves\-tor} badge.

Second, the introduction of the \badge{Tag editor} and \badge{In\-ves\-tor} badges was followed by an increase on the average intensity of the corresponding first-time 
action of more than $4$$\times$, from $\hat{k}_0 \leq 2 \cdot 10^{-4}$ to $\hat{k}_1 \approx 8 \cdot 10^{-4}$ for \badge{Tag editor} badge and from $\hat{k}_0 \leq 5 \cdot 10^{-5}$ 
to $\hat{k}_1 \approx 2 \cdot 10^{-4}$ for \badge{In\-ves\-tor} badge. 
Equivalently, the average time a user takes to perform the actions for the first time was reduced by 75\%. 
Moreover, this change in user behavior did not vanish over time (rightmost column).

Finally, in the case of the \emph{In\-ves\-tor} badge, we find a transient increase on the average intensity of bounties to other users'{} questions around October-November 2010, 
which is statistically significant. 
Upon investigation, we notice that several users discovered ways of benefiting from offering bounties around that time, triggering subsequent first-time uses of bounties by other 
users\footnote{\scriptsize https://meta.stackexchange.com/questions/64824/clever-bounty-reputation-hack}. Such discussions led to an increase on the minimum reputation one can 
transfer when offering a bounty.

\begin{table}[!t]
\caption{Absolute Standardized Mean Difference (|SMD|) between the treatment and the control groups. 
} 
\small
\centering
\begin{tabular}{ l | c c |  c c } 
&
\multicolumn{2}{c}{\badge{Tag Editor}} &
\multicolumn{2}{c}{\badge{Prom. \& Inv.}} \\
 \hline	
feature & {\small $\bar{\text{|SMD|}}$} & {\small $\sigma_{\text{|SMD|}}$} & {\small $\bar{\text{|SMD|}}$} & {\small $\sigma_{\text{|SMD|}}$}  \\
  \hline	
age & 0.10 & 0.04  & 0.14 & 0.10\\
age-NA$^{*}$ & 0.08 & 0.12  & 0.17 & 0.11\\
reputation & 0.21 & 0.35  & 0.14 & 0.11\\
\#views & 0.16 & 0.23   & 0.11 & 0.08\\
\#upvotes & 0.18 & 0.28 & 0.13 & 0.10 \\
\#downvotes & 0.08 & 0.07  & 0.03 & 0.02 \\
 \hline  
\end{tabular}
{\tiny
\\$^{*}$Balance check on a binary variable indicating if the value is missing. 
}
\label{tab:smd}
\vspace{-2mm}
\end{table}

\section{Badges and utilities}
\label{sec:utility}

In this section, we look for plausible reasons that explain
\emph{why} the \badge{Tag editor} and \badge{In\-ves\-tor} badges were successful at steering users'{} behavior while the \badge{Pro\-mot\-er} 
badge was unsuccessful.
To this aim, we resort to game-theoretic concepts such as user utilities, action payoffs and reservation values, and identify 
measurable proxies of some of these concepts.

\xhdr{User utilities}
In the game theory literature~\cite{easley2016incentives, ImmStoSyr15, zhang2016social}, the utility a user obtains from performing an action is defined 
as the difference between the action payoff $p$ and the cost of effort $c$, \ie, $v = p - c$.
Moreover, the fact that participation is a voluntary, strategic choice---users have a choice about whether to perform an action---is often modeled 
via a reservation value $\omega$ that the utility $v$ must exceed in order for the user to perform the action. 
More specifically, if $p - c < w$, the user will decide not to perform the action and, otherwise, she will perform it.
In this context, a badge $b$ is assumed to increase the utility a user obtains from performing the action, \ie, $v = p - c + v_b$, where $v_b$ is the badge 
\emph{value}. 
Then, depending on the actual values of $p$, $c$, $v_b$ and $\omega$, one can argue that a badge will induce users to perform an action that, in the absence of a badge, would not perform.

However, in social media sites and online communities, the action payoffs, cost of effort, badge value and reservation values are typically intangible, hidden or ambiguously defined.
As a consequence, our causal inference framework did not explicitly adopt 
 the above model and instead used a data-driven approach based on survival analysis 
 using only the observable temporal traces.
In this section, however, 
we turn our attention towards the above stylized model, try to identify measurable proxies of the model parameters for each of the studied badges and 
actions and use them to investigate the plausible reasons for the success or failure of badges at steering users'{} behavior, as concluded by our framework.

\xhdr{Proxies to user utilities} 
We consider the following observable proxies for the utilities users obtain from editing a wiki tag and offering a bounty, respectively:
\begin{itemize}
\item[(a)] \emph{Tag popularity}: 
the more popular a tag is, the greater the utility $v$ a user may obtain from editing its wiki tag---the user needs to put less effort to
create a wiki on a popular tag and she receives the satisfaction of helping a larger part of the community. 

\item[(b)] \emph{Number of answers}:
the higher (lower) the number of answers a question receives after (before) offering a bounty, the greater the utility $v$ a user obtains 
from offering the bounty. 
Moreover, users offering a bounty for an answer to another user'{}s question may obtain less utility from the answers since they did not 
originally ask the question.
Here, note that we did not find a correlation between the bounty value---its cost---and the final number of answers ($\rho \approx 0.1$)
and thus we can safely ignore the bounty value.

\end{itemize}

Given the above proxies, we proceeds as follows. In terms of tag wikis, we group tags by popularity (\ie, number of questions a tag was used on) 
and model the time the users take to create a wiki for a tag of a given popularity $p$ 
as a survival process. Moreover, we characterize this process using an intensity $\lambda_p(t)$, which we define as follows:
\begin{equation} \label{eq:intensity-popularity}
\lambda_p(t) =
\begin{cases}
  0 & \text{if}\enskip t < s  \\
  {\lambda_0(p)} & \text{if}\enskip s  \leq t < \tau\\
  {\lambda_1(p)} & \text{otherwise}
\end{cases}
\end{equation}
where $\lambda_0(p)$ and $\lambda_0(p)$ are parameters shared across all tags with popularity $p$, $s$ is the time when the tag  is first used in a
question, and $\tau$ is the time when the \badge{Tag Editor} badge is introduced. 
Then, by comparing the maximum likelihood estimators of the model parameters, $\hat{\lambda}_0(p)$ and $\hat{\lambda}_1(p)$, for different popularity levels $p$, 
we can assess the causal effect of the \emph{Tag Editor} badge on tag wikis with different utility values.

In terms of bounties, we first group questions by the number of answers they received in the first two days since they were asked and then compare the additional number 
of answers they received after those first two days if a bounty was (not) offered in the second day.
Moreover, for questions that received a bounty, we estimate the distribution of the number of answers they received before the bounty was offered both before and after the badges \badge{Pro\-mot\-er} and \badge{In\-ves\-tor} were introduced.
By controlling for the number of answers, we can assess the causal effect of both badges for bounties with different utility values.

\begin{figure}[t]
\vspace{1mm}
\centering
\includegraphics[width=0.315\textwidth]{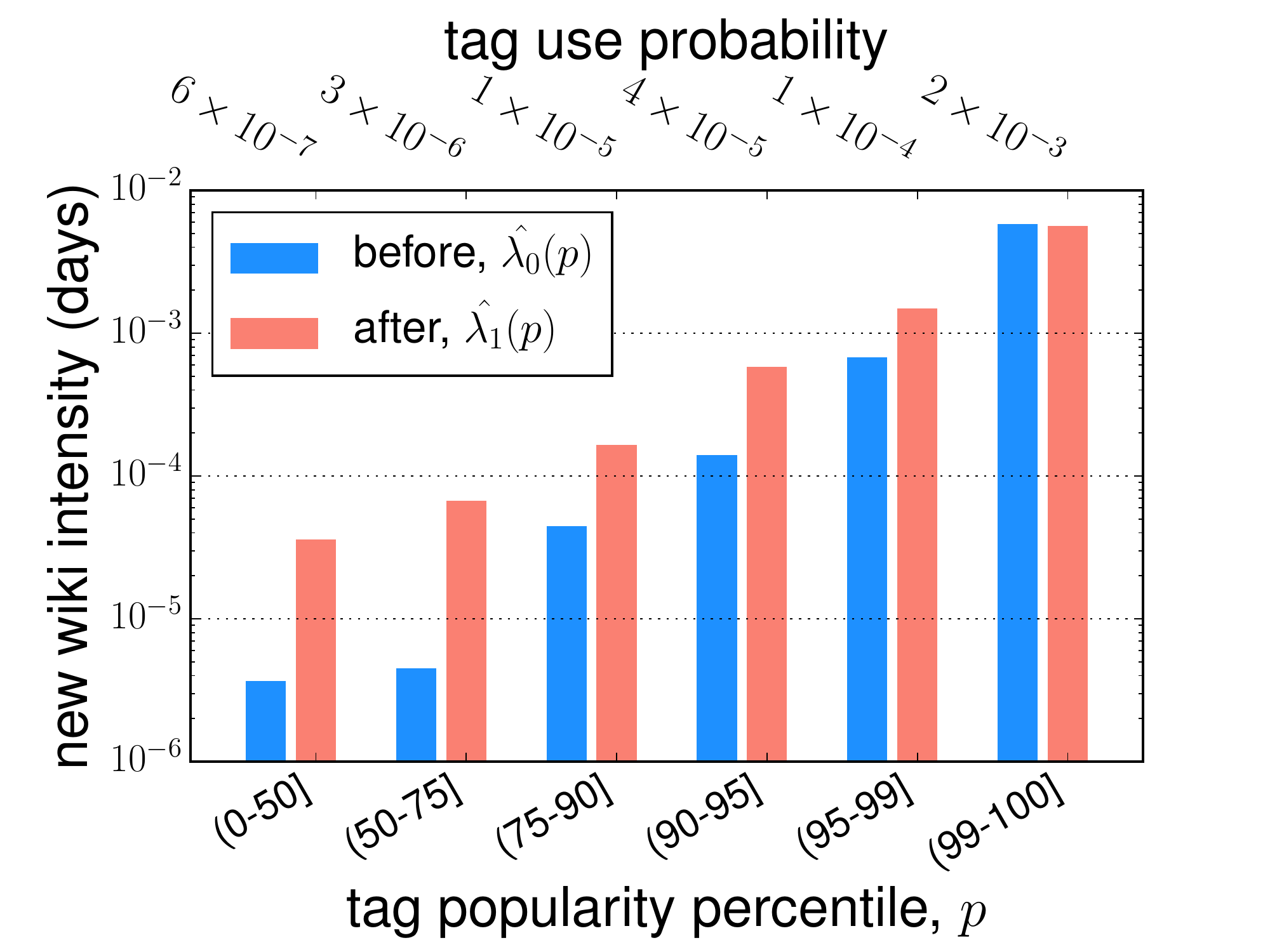}
\vspace{-1mm}
\caption{Causal effect of the \badge{Tag Editor} badge for tags with different utility value, as estimated by their
popularity level.}
\label{fig:tagedits_utility}
\vspace{-3mm}
\end{figure}

\xhdr{Results}
Figure~\ref{fig:tagedits_utility} summarizes the results for the \badge{Tag Editor} badge, which shows the higher the popularity (utility) of a tag, 
the weaker the causal effect of the badge introduction.
For example, while the intensity of the tags at the bottom 50\% in terms of popularity increased an order of magnitude after the badge was introduced, 
the intensity of the tags at the top 1\% did not change.
This suggests that the introduction of the badge steered users to create tag wikis for less popular, low utility tags.

\begin{figure}[t]
\vspace{-3mm}
\centering
\captionsetup[subfigure]{labelformat=empty, justification=centering}
\subfloat[Total number of answers vs number of answers preceding bounty offering]{
\includegraphics[width=0.315\textwidth]{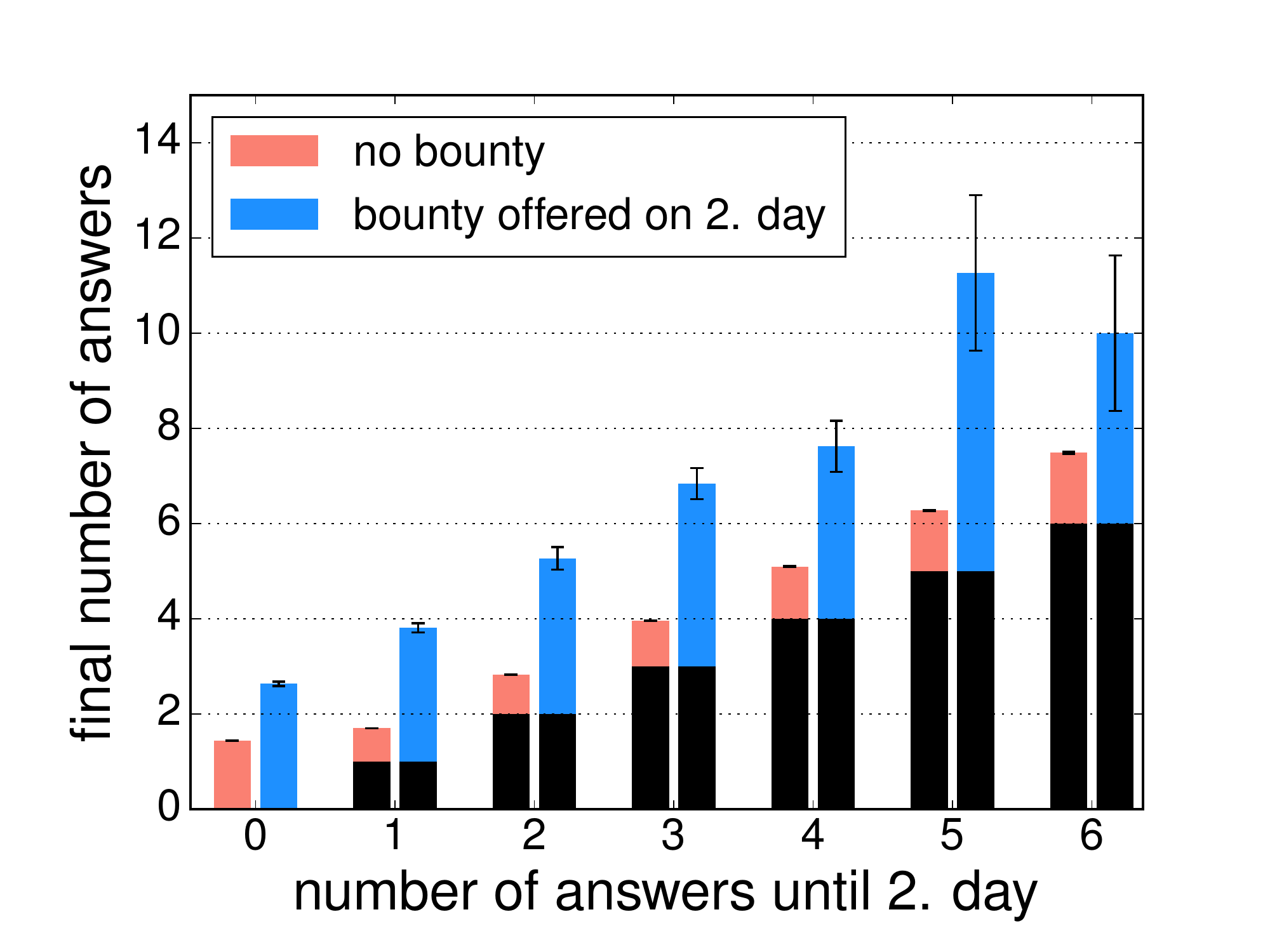}
\label{fig:bounties:increase}
} 
\subfloat[Number of answers preceding bounty offering]{
\includegraphics[width=0.315\textwidth]{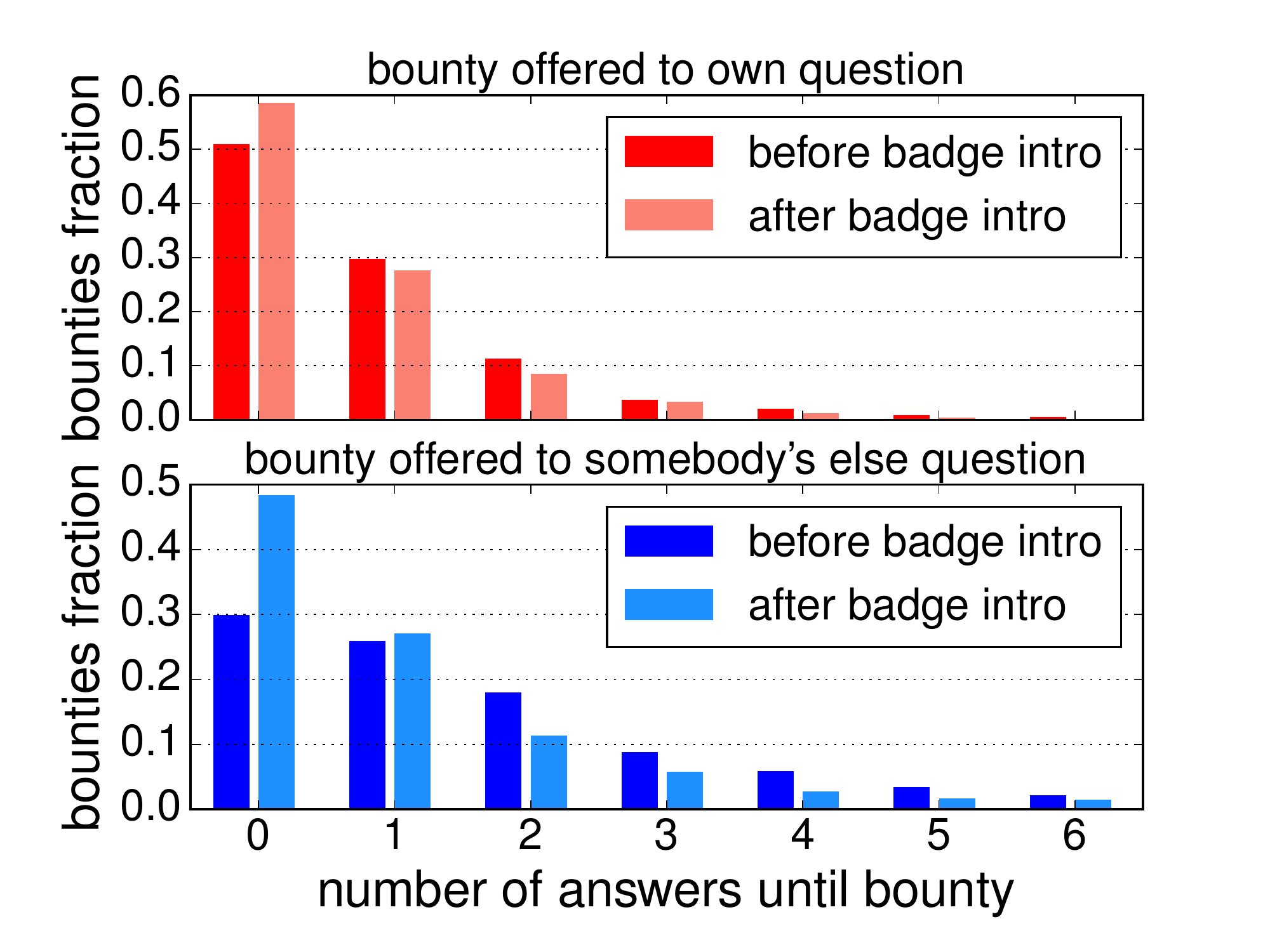}
\label{fig:bounties:change}
}
\vspace{-1mm}
\caption{Causal effect of the \badge{Pro\-mot\-er} and \badge{In\-ves\-tor} badges for bounties with different utility value, as estimated by the number of answers 
preceding the bounty offering.}
\label{fig:bounties_utility}
\vspace{-3mm}
\end{figure}

Figure~\ref{fig:bounties_utility} summarizes the results for the \badge{Pro\-mot\-er} and \badge{In\-ves\-tor} badges, which let us better understand their failure 
and success, respectively:
(i) the number of answers a bounty \emph{triggers} (\ie, its utility) increases with the number of answers the question has received in its absence (top panel); 
(ii) the introduction of the \badge{Pro\-mot\-er} badge changed the users'{} willingness to offer bounties to their own questions only up to a very small degree 
(top panel, top figure), whereas the \badge{In\-ves\-tor} badge did change it remarkably for bounties offered to other users'{} questions (bottom panel, bottom figure). 
In both cases, the change is statistically significant, \ie, 
$\chi^2=86.2$, $p<0.001$ for \badge{Pro\-mot\-er} and $\chi^2=114.9$, $p<0.001$ for \badge{In\-ves\-tor} badge using Mood's median test,
however, only the latter was sufficiently pointed to result in a significant change at an individual user level, as concluded in Section~\ref{sec:results}.

\xhdr{Remarks}
For the specific actions and badges under studied, we were able to manually identify sensible proxies of the users'{} utilities, however, 
finding sensible proxies for other types of actions in Stack Overflow or other online platforms will require careful reasoning and justification. 

\section{Do badges improve the community functioning?}
\label{sec:community}
In this section, we use a survival-based counterfactual analysis, in which we investigate \emph{what would have happened} if the
\badge{Tag Editor} and \badge{In\-ves\-tor} badges had not been introduced, to assess to which extent the badges improved the site 
functioning at a community level.

\xhdr{Counterfactual analysis}
For the \badge{Tag editor}, we assess the site functioning at a community level in terms of the number of new tag wikis over time 
and, for the \badge{In\-ves\-tor} badges, we use the time to bounty and time to first answer across questions. More specifically, 
we proceed as follows.

\emph{--- New tag wikis}: 
we simulate the time the users take to create a new wiki for a tag of a given popularity $p$ in the counterfactual world where the \badge{Tag editor} 
badge is never introduced using the intensity defined by Eq.~\ref{eq:intensity-popularity} with $\lambda_{0}(p) = \lambda_{1}(p) = \hat{\lambda}_0(p)$, 
where $\hat{\lambda}_0(p)$ is the maximum likelihood estimate for $\lambda_{0}(p)$ in the true world.
Then, we compare the number of new tag wikis over time as well as the popularity of their associated wikis in the true world and in the simulated 
counterfactual world.

\emph{--- Time to bounty and first answer}:
for questions that received a bounty, we model the time that the users take to offer the bounty as a survival process with associated intensity $\lambda_b(t)$,
which we define as follows:
\begin{equation} \label{eq:intensity-bounty}
\lambda_b(t) =
\begin{cases}
  0 & \text{if}\enskip t < s \\
  {\lambda_0(b)} & \text{if}\enskip s \leq t < \tau\\
  {\lambda_1(b)} & \text{otherwise}
\end{cases}
\end{equation}
where the parameter $b \in \{0, 1\}$ denotes whether the badge is offered by the user asking the question or by another user, $\{\lambda_i(b)\}_{i, b \in \{0, 1\}}$ are (four) parameters shared 
across all questions, $s$ is the time since two days after the question is asked\footnote{\scriptsize A bounty can only be offered two days after the question has been asked.}, and $\tau$ is the time 
when the \badge{Pro\-mot\-er} and \badge{In\-ves\-tor} badges are introduced.
For all questions, we model the time that the users take to provide the first answer also as a survival process with an associated intensity defined similarly as in Eq.~\ref{eq:intensity-bounty}, 
however, in this case, the parameter $b \in \{-1, 0, 1\}$ denotes whether the question received a badge and, if so, whether it was offered by the user asking the question or by another user,
and thus the model has six parameters.
Then, we compare the maximum likelihood estimators of the above parameters to assess what would have happened if the \badge{In\-ves\-tor} badge had not been introduced.

\begin{figure}[t]
\centering
\captionsetup[subfigure]{labelformat=empty}
\captionsetup[subfigure]{justification=centering}
\subfloat[Number of tags over time]{
\includegraphics[width=0.315\textwidth]{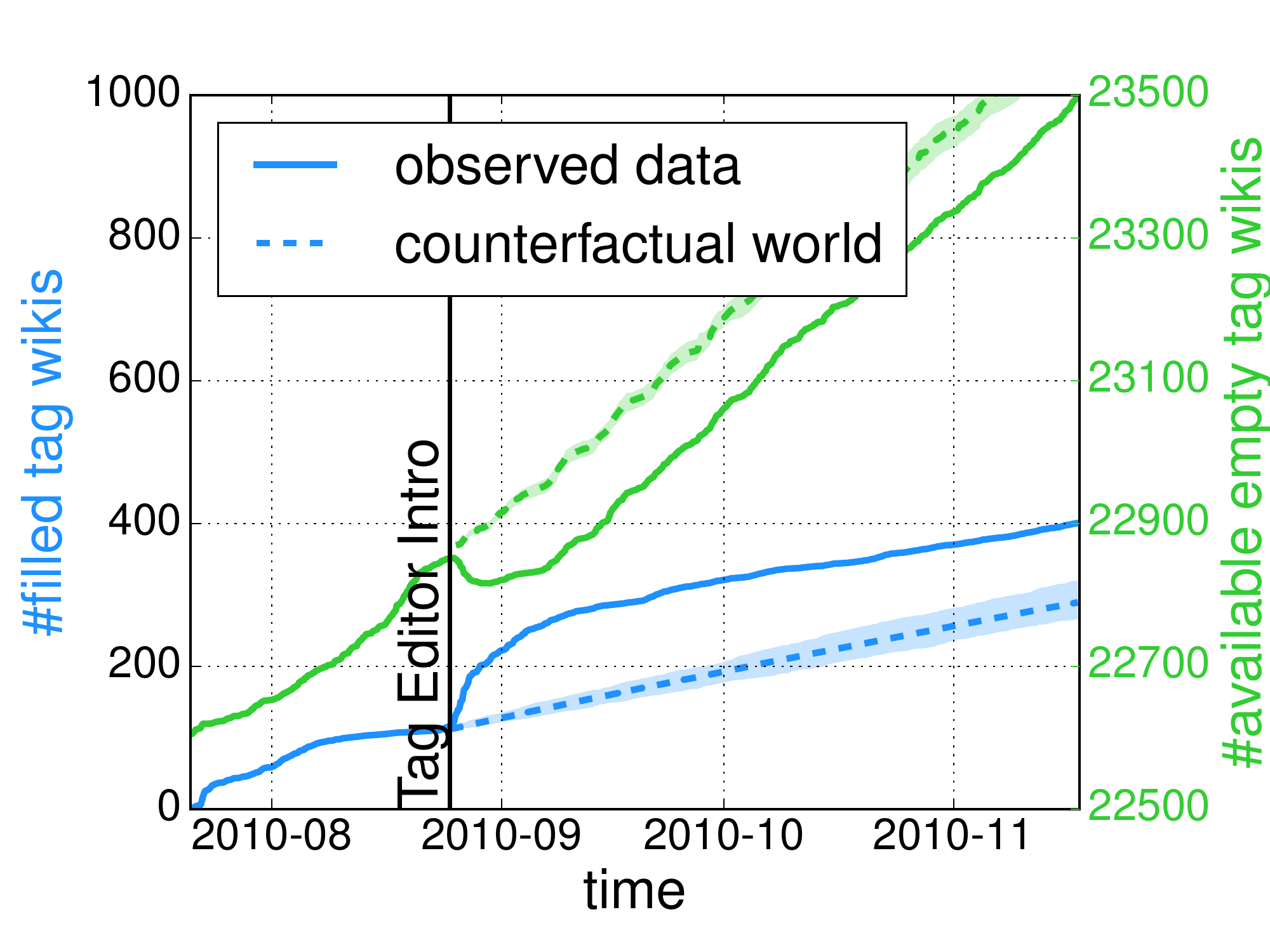}} 
\subfloat[Tag wiki rank over time]{
\includegraphics[width=0.315\textwidth]{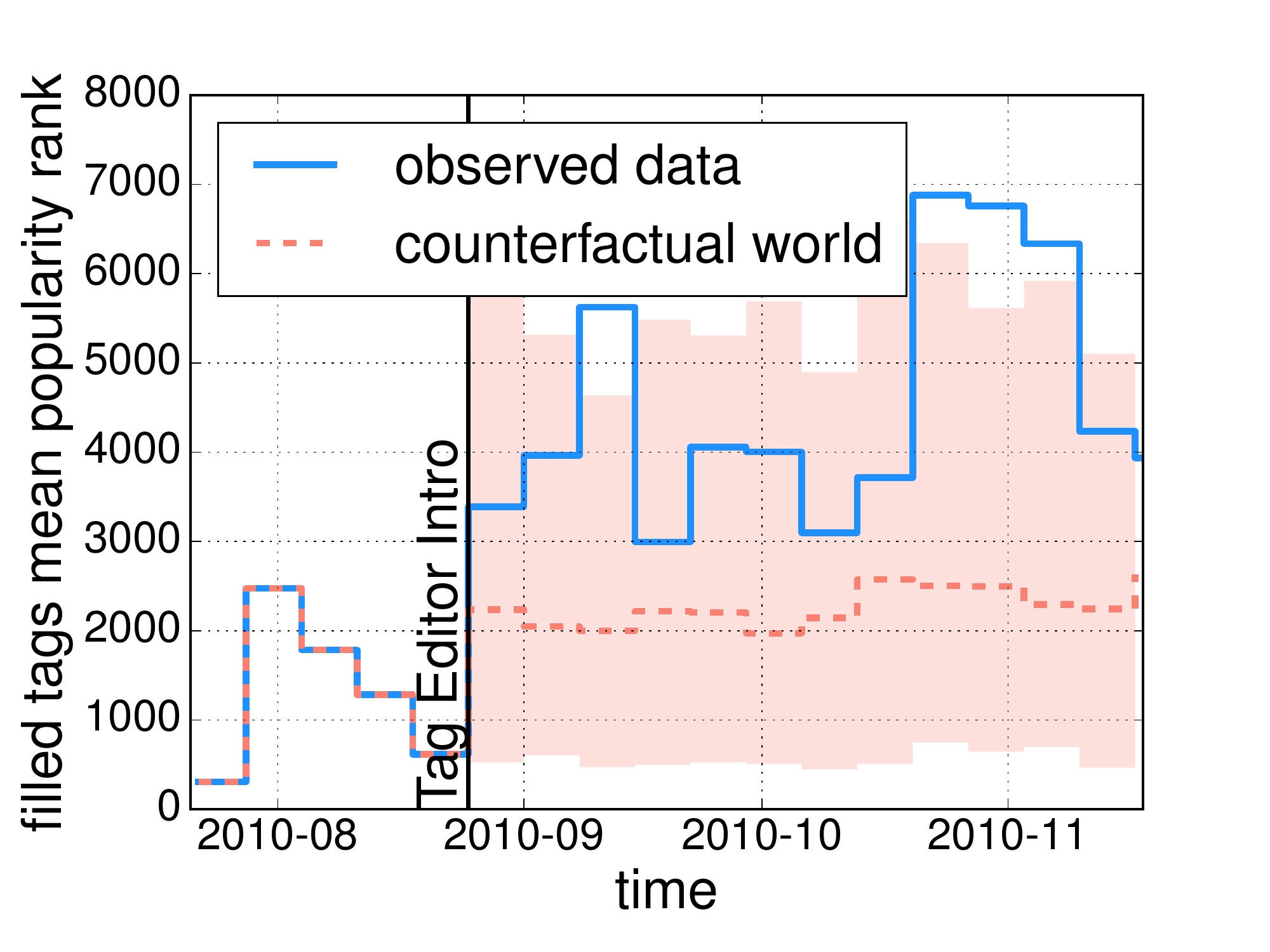}}
\vspace{-1mm}
\caption{Tag wikis with and without the \badge{Tag Editor} badge. Simulations means and 95\%-CI shown. }
\label{fig:tagedits_benefit}
\vspace{-2mm}
\end{figure}

\xhdr{Results}
Figure~\ref{fig:tagedits_benefit} summarizes the results for the \badge{Tag Editor} badge. 
The top panel shows the number of unique tags with (blue) and without (green) supporting tag wiki over time in the true and
in the simulated counterfactual world. Note that the number of unique tags without tag wiki grows over time as users ask questions 
with new (not previously used) tags.
The results show that the tags with supporting tag wikis increase at a higher rate in the true world than in the counterfactual world 
just after the badge introduction, however, after a relatively short period of time, the rate decreased to its original value and match 
the rate at which tags with tag wikis grows in the counterfactual world.
The bottom panel, which shows a decrease on the average rank popularity (rank) of the tags with supporting tag wikis, supports the following 
hypothesis to explain this phenomenon.
The badge lifted the utility value of editing tag wikis, however, this increase was large enough to exceed the reservation value only for 
tags of certain popularity. As a consequence, after these tags had a tag wiki, the effect of the badge diminished.

\begin{figure}[t]
\centering
\vspace*{1mm}
\captionsetup[subfigure]{labelformat=empty}
\captionsetup[subfigure]{justification=centering}
\subfloat[Time to bounty]{
\hspace*{-2mm}
\includegraphics[width=0.315\textwidth]{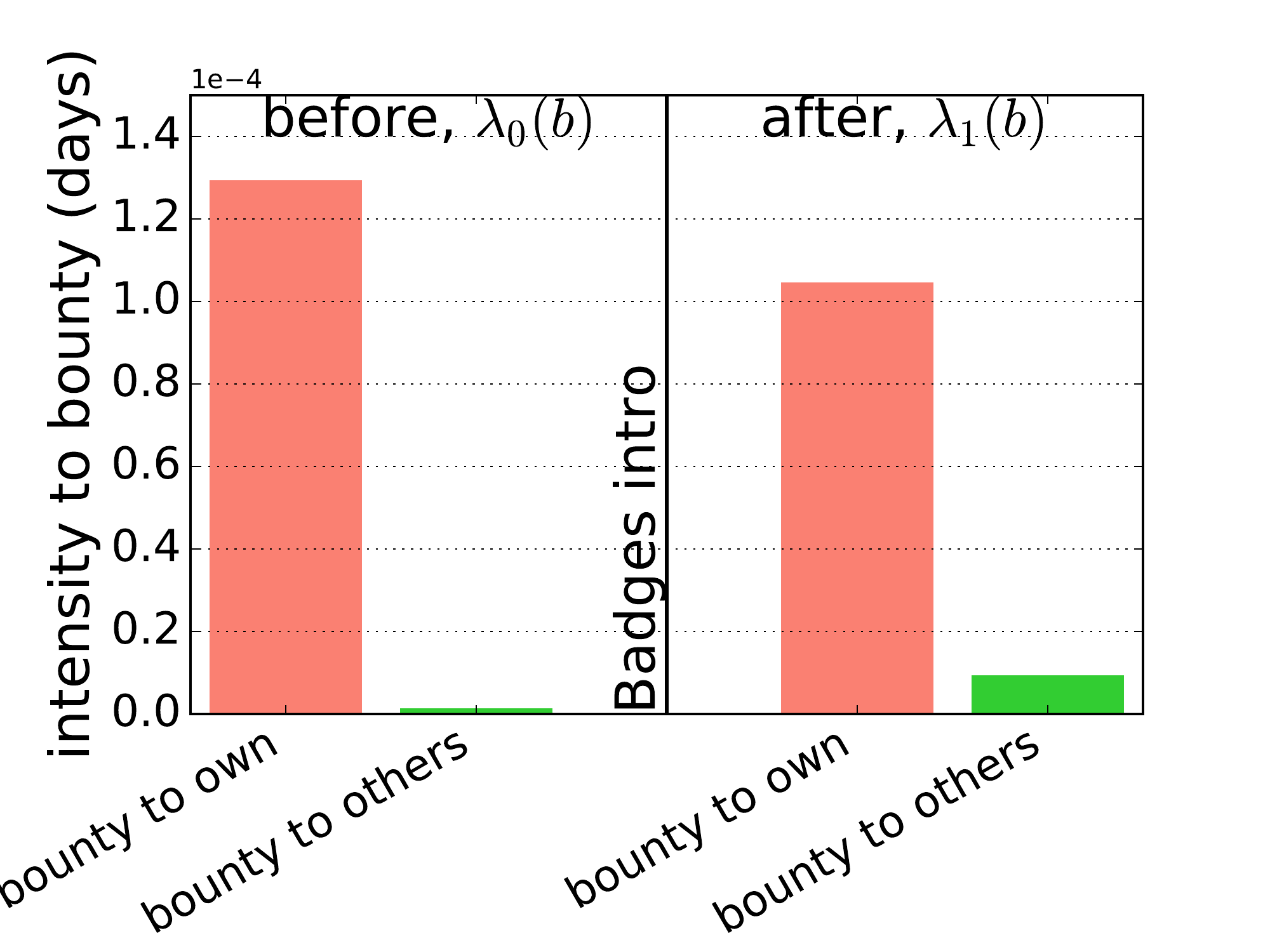}
}
\subfloat[Time to first answer]{
\hspace*{-5mm}
\includegraphics[width=0.315\textwidth]{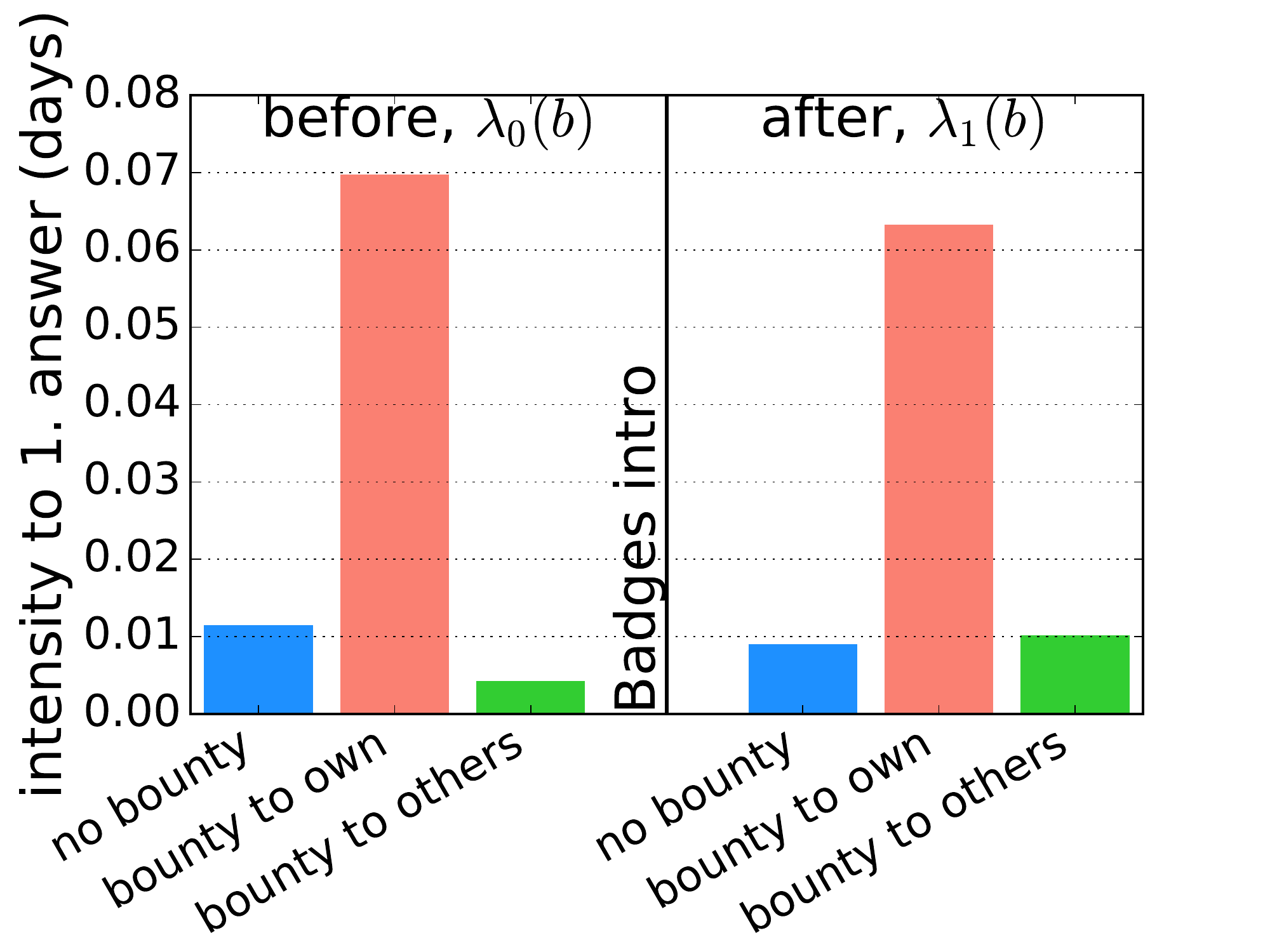}
}
\vspace{-2mm}
\caption{Time to bounty and first answer with and without the \badge{Pro\-mot\-er} and \badge{In\-ves\-tor} badges.}
\label{fig:bounties_benefit}
\vspace*{-3mm}
\end{figure}

Figure~\ref{fig:bounties_benefit} summarizes the results for the \badge{In\-ves\-tor} badge, which show that the time to bounty and first answer for questions 
in which a bounty was offered by a user different than the user asking the question decreased (\ie, the intensities increased) after 
the \badge{In\-ves\-tor} was introduced, improving the site functioning.
In contrast, the time to first answer (and time to bounty) in questions without bounty (and with a bounty offered by the user asking the question) increased.
The latter observation suggests that the \badge{In\-ves\-tor} badge may have mitigated the global slowdown in the time users take to answer questions by 
motivating more people to offer bounties faster---in the counterfactual world where the \badge{In\-ves\-tor} badge had not been introduced, the site functioning 
would have actually worsened.

\section{Conclusions}
\label{sec:conclusions}
Social media sites and online communities are dynamic environments where users change their behavior on a daily basis due to
many complex factors. As a consequence, assessing the effectiveness of incentive mechanisms, which are ubiquitous among them, 
is challenging. 
In this work, we have focused on one of the simplest incentive mechanisms---first-time badges---and studied their effectiveness by 
developing a novel survival-based causal modeling framework, specially designed to harness the delayed introduction of several 
badges in a popular Q\&A website.

Our work also opens up many interesting venues for future work. For example, it would be very interesting to use our framework to 
analyze badges in other online platforms as well as extend it to other types of badges, \eg, threshold badges.
Badges are typically awarded to all users whose contributions \emph{exceed} some predefined values, however, it would be interesting to also 
consider incentive mechanisms where users compete for a limited reward. 
Finally, in this work, we have focused on assessing the causal effect of (first-time) badges. A natural follow-up would be developing principled, effective 
methods to optimize their design.

\section*{Acknowledgements} 
We thank Utkarsh Upadhyay and Isabel Valera for useful discussions.

{
\bibliographystyle{abbrv}
\bibliography{refs}
}

\end{document}